\documentclass[preprint,aps,showpacs,preprintnumbers,amsmath,amssymb,nofootinbib]{revtex4}
\usepackage{epstopdf}
\usepackage{epsfig}
\usepackage{graphicx}

\newcommand{\be}{\begin{equation}}
\newcommand{\ee}{\end{equation}}
\newcommand{\bea}{\begin{eqnarray}}
\newcommand{\eea}{\end{eqnarray}}

\usepackage{float}

\begin{document}
\title{\large Non-unitary lepton mixing in an inverse seesaw and its impact on the physics potential of long-baseline experiments}

\author{Soumya C. and Rukmani Mohanta }
\affiliation{School of Physics, University of Hyderabad, Hyderabad - 500 046, India }

\begin{abstract}
In this paper, we consider the low-energy scale inverse seesaw mechanism in which the observed neutrino mass and lepton mixing are explained by introducing right handed neutrinos and the gauge-singlet
fermions with experimentally testable energy scale. Moreover, the presence of such new fermions leads to unitarity violation in lepton mixing due to significantly large mixing between active neutrinos and the heavy fermions. In addition to this, such large lepton mixing also gives rise to potentially large lepton flavor violation, which allows to constrain the non-unitarity parameters via lepton flavor violating decays ($l_i \to l_j \gamma$). We make use of these constraints on non-unitarity parameters and investigate their effects on the determination of current unknown oscillation parameters at long-baseline experiments. We find that  non-unitarity parameters are sensitive to NO$\nu$A experiment. However, it is observed that NO$\nu$A experiment is not expected to improve the current knowledge of non-unitarity parameter $\eta_{21}$.  We also find that the sensitivities to current unknowns are deteriorated significantly in presence of non-unitary lepton mixing and these sensitivities crucially depend upon the new CP-violating phase in the non-unitary mixing. Further, we find that the degeneracy resolution capability of NO$\nu$A experiment is reduced in the presence of non-unitarity parameters. However, the synergy between the currently running experiments T2K and NO$\nu$A can improve the parameter degeneracy resolution and hence there is enhancement in the sensitivities of unknowns.

\end{abstract}

\maketitle
\section{Introduction}
The confirmation of neutrino oscillation by atmospheric, solar, reactor, and accelerator neutrino oscillation experiments \cite{exp-1,exp-2,exp-3,exp-4,exp-5,exp-6,exp-7,exp-8} has been the first ever evidence for New Physics (NP) beyond the Standard Model (SM). So far huge progress has been made in extracting the information about the  knowledge of the neutrino masses and lepton flavor mixing parameters. Moreover, the three flavor neutrino oscillation has become the standard picture of neutrino flavor transitions. However, the short-baseline anomalies \cite{Aguilar:2001ty, AguilarArevalo:2008rc, Mention:2011rk, Mueller:2011nm, Aguilar-Arevalo:2012fmn}  hint towards existence of extra one or more  neutrino states, so-called sterile neutrinos. Such neutrino states are present in plenty of neutrino mass models and their mass scale can vary from well below the electroweak scale upto the Plank scale. Apart from these neutrino mass models, the existence of sterile neutrinos are also motivated by various cosmological observations \cite{Dvorkin:2014lea, Hamann:2013iba, Wyman:2013lza, Battye:2013xqa}. Consequently, theoretically and experimentally motivated sterile neutrino has become the smoking-gun signal for the New Physics beyond the standard paradigm of neutrino flavor transition.

If sterile neutrinos exist in nature, then in principle they can mix with active neutrinos  which results in unitarity violation in the active neutrino mixing matrix (PMNS matrix). Therefore, any deviation from the unitarity of PMNS matrix  points toward presence of sterile neutrinos. As the light sterile neutrinos can be produced  at neutrino oscillation experiments, they can  be probed via neutrino oscillation physics. Whereas, the production or detection process of heavy sterile neutrinos are kinematically forbidden in low energy neutrino experiments and such sterile neutrinos can be probed by looking at the  deviations in the unitarity  of lepton mixing matrix. As of now, numerous  experiments are  probing the existence of sterile neutrinos, but  none of such particles have been detected so far. For instance, search for light sterile neutrinos with the IceCube detector has found no evidence for their existence \cite{TheIceCube:2016oqi}. Moreover, the recent results from NO$\nu$A experiment also could not see any signal for the existence of light sterile neutrino \cite{Adamson:2017zcg}. Therefore,  in this work, we examine whether the non-unitarity effects, which are arising from the mixing between active neutrino and heavy sterile neutrinos present in a low-scale seesaw model, can be probed at long-baseline experiments.

The formulation of natural and viable mechanism to accommodate neutrino mass in the SM is a challenging task in the theoretical point of view. The seesaw mechanisms (Type I \cite{Mohapatra:1979ia}, II \cite{Magg:1980ut, Schechter:1980gr, Lazarides:1980nt, Mohapatra:1980yp, Wetterich:1981bx}, and III \cite{Ma:1998dn, Ma:2002pf, Hambye:2003rt}) are the most captivating  theoretical frameworks, which could explain the lightness of neutrino mass by the introduction of heavy new particles. The main drawback of these models is that the  energy scale of the new particles is  approximately $10^{14}~ \rm{GeV}$ (GUT scale) and therefore, these particles are out of reach of current or even future collider experiments. In contrast to this, the low-energy seesaw mechanism like inverse seesaw mechanism \cite{Mohapatra:1986bd} gets more attention since  the energy scale of the new particles in this model is of the order of $\rm{TeV}$ scale and hence, it can be experimentally testable. As the mixing  between the active and sterile neutrinos inversely related to the new physics scale i.e, ${\cal O} \left ( \frac{m_D}{M_N}\right )$ ($M_N$ is mass of new particle), such mixing is quite large in the inverse seesaw mechanism unlike canonical seesaw mechanism. Therefore, in low-scale seesaw model the deviation from unitarity of lepton mixing matrix is significantly large and this is the reason why we are focusing on inverse seesaw.

The various aspects of non-unitary lepton mixing are extensively discussed in the literature in both phenomenological and theoretical perspectives \cite{Antusch:2006vwa,Goswami:2008mi, Dev:2009aw, Abada:2012mc, Abada:2013aba, Awasthi:2013ff, Abada:2014kba, Emelyanov:2014jna, Antusch:2016brq, Escrihuela:2015wra, Blennow:2016jkn}. In \cite{Malinsky:2009df}, it has been  shown that neutrino factory experiment  can provide an excellent probe for non-unitarity effects which are  emerging in the minimal inverse seesaw model. Furthermore, there are studies which dedicated to constrain the non-unitarity parameters \cite{Antusch:2008tz,Antusch:2014woa}. Some of the recent studies which have discussed the consequences of non-unitarity effect  on the determination of neutrino mass hierarchy, octant of atmospheric mixing angle, and CP violating phase by long-baseline experiments can be found in \cite{Escrihuela:2016ube, Dutta:2016czj, Dutta:2016vcc, Verma:2016nfi}. In this paper, we focus on  low energy scale inverse seesaw model which permits significantly large mixing between the active and sterile neutrinos and gives rise to non-unitary lepton mixing.  Moreover, the constraints on the non-unitarity parameters can be obtained from  the lepton flavor violating decays ($l_i \to l_j \gamma$) which are mediated by the heavy particles present in the inverse seesaw model \cite{Forero:2011pc}. We make use of these constraints on non-unitarity parameters and investigate their effects on the determination of current unknowns in oscillation sector by long-baseline experiments.

The paper is organized as follows. 
We review non-unitary lepton mixing in an inverse seesaw model in section  II. The effect of non-unitary mixing on neutrino oscillation and  its implications at long-baseline experiments are respectively discussed in sections III and IV. Finally, we conclude in section V.
%
\section{Non-unitary mixing in an inverse seesaw}
The origin of the observed neutrino masses is one of the great open questions in particle physics. Among the various theoretical attempts to explain lightness of neutrino mass, the  low-scale seesaw models are the well accepted ones, because the new particles in these models are within the reach of collider experiments. We briefly describe below the low-scale inverse seesaw mechanism, which can provide considerable non-unitarity effects.

The low-scale seesaw model is constructed by extending the Standard Model (SM) particle content with right-handed neutrinos ($\nu_{R}$) and sterile fermions ($S_L$) and assuming a $U(1)_L$ global lepton number symmetry \cite{Mohapatra:1986bd, jwfvalle}, with the lepton  number of right-handed neutrinos and sterile fermions are chosen to be $+1$ and $-1$ respectively, and the neutrinos get masses only when $U(1)_L$ symmetry is broken.  Thus, one can write the effective Lagrangian for neutrino mass in presence of these new particles, which is of the form
\begin{equation}
-\mathcal{L}= \bar{l}_LY_{\nu}\Phi^c \nu_{R}  + \frac{1}{2} \bar{\nu}_R^c M_R \nu_R + \bar{\nu}_R^c M S_L + \frac{1}{2} \bar{S}^c_L \mu S_L +{\rm h.c.},
\end{equation}
where $l_L$ and $\Phi$ are  lepton and Higgs doublets in the SM, $Y_{\nu}$  is the Yukawa coupling matrix and  $M_R$ and $\mu$ respectively are the  Majorana mass matrices for right-handed neutrino and sterile fermion, which are symmetric in nature. The spontaneous symmetry breaking in Higgs sector yields
\begin{equation}
-\mathcal{L}= \bar{\nu}_L M_D \nu_{R}  + \frac{1}{2} \bar{\nu}_R^c M_R \nu_R + \bar{\nu}_R^c M S_L + \frac{1}{2} \bar{S}^c_L \mu S_L +{\rm h.c.},
\end{equation}
where $M_D = Y_{\nu} \langle \Phi\rangle $. The above Lagrangian can be expressed in a mass matrix form as
\begin{equation}
-\mathcal{L} = \frac{1}{2} \left(\begin{array}{ccc}
\bar{\nu}_L & \bar{\nu}_R^c & \bar{S}^c
\end{array} \right) 
\left( \begin{array}{ccc}
0&M_D& 0\\
M_D^T&M_R&M\\
0&M^T&\mu \\
\end{array} \right)
 \left( \begin{array}{c}
 \nu_L^c \\ \nu_R \\ S\\
 \end{array} \right) +{\rm h.c.}.
\end{equation}
If one assumes that lepton number is violated only in Majorana mass terms of sterile fermion, i.e.,  $\mu$ is non zero and $M_R =0$.  As a result, one ends up with neutrino mass matrix for an inverse seesaw model and it is given by
\begin{equation}
\mathcal{M}_{\nu}=
\left( \begin{array}{ccc}
0&M_D&0\\
M_D^T&0&M\\
0&M^T&\mu\\
\end{array}\right).
\end{equation}
It is appropriate to consider the Dirac mass ($M_D$) of neutrino of the order of GeV scale (electroweak scale). As the order of lepton number violation in nature is too small, the  $\mu$ parameter is considered to be small, i.e.,  $\mu \approx \text{keV}$. Moreover, $M$ is a SM singlet mass term, which is not governed by the $SU(2)_L$ symmetry breaking.  Therefore,    one can consider  $\mu \ll M_D<M$ with  $M$ is of the order of \rm{TeV} scale. With these assumptions, one can block diagonalise $\mathcal{M}_{\nu}$ into heavy and light sectors, which yields the light neutrino mass matrix (so-called inverse seesaw formula) as
\begin{equation}
m_{\nu} =M_DM^{-1}\mu (M^T)^{-1}M_D^T =F \mu F^T, \label{iss}
\end{equation} 
where $F=M_DM^{-1}$. It can be inferred from the above equation that for $\mu$   of the order of \rm{keV} scale, $F\approx10^{-2}$  leads to desired sub-eV scale neutrino masses. Further, the diagonalization of $m_{\nu}$ yields the light neutrino mass as
\begin{equation}
U_\text{PMNS}^{\dagger}m_{\nu}U_\text{PMNS}^* = \text{diag}(m_1,m_2,m_3).
\end{equation} 
The mass matrix $\mathcal{M}_{\nu}$  can be diagonalised by an unitary matrix $U_{eff}$, which yields the mass matrix in the mass basis as
\begin{equation}
U_{eff}^{\dagger}\mathcal{M}_{\nu}U_{eff}^* =\tilde{m}_i=\text{diag}(m_{\nu_i},m_{s_j})\;,
\end{equation}
where $m_{\nu_i}$ ($i=1,2,3$) are the Majorana light neutrino masses and $m_{s_j}$ ($j=4,5, \cdots, 9$) are the  pseudo Dirac neutrino masses. Further, the effective unitary mixing matrix is of the form
\begin{equation}
U_{eff} =\left( \begin{array}{cc} N_{3 \times 3} & \Theta_{3 \times 6} \\ R_{6 \times 3} & S_{6 \times 6}\\ 
\end{array} \right),
\end{equation}
where $N_{3 \times 3}$ is the non-unitary active neutrino mixing matrix, which can be parametrized as
\begin{equation}\label{eta}
N =(1-\eta)U_\text{PMNS},
\end{equation}
$\Theta_{3 \times 6}$ contains the light-heavy mixing elements, $R_{6 \times 3}$ corresponds to heavy-light  mixing elements, and $S_{6 \times 6}$ corresponds to heavy-heavy mixing elements. 
Thus,  the mixing matrix which is used to diagonalise the light neutrino mass matrix ($m_{\nu}$) is no more unitary and is given by \cite{Schechter:1981cv}
\begin{equation}\label{mdl}
N=(1-\frac{1}{2}\Theta^{\dagger}\Theta)U_\text{PMNS},
\end{equation}
which yields,
\begin{equation}
\eta = \frac{1}{2}\Theta^{\dagger}\Theta.
\end{equation}
In order to find the non-unitarity parameters, one can make use of Casas-Ibarra parametrization for $\Theta$ as discussed in \cite{Ibarra:2003up}, which is given by
\begin{equation}
\Theta =U_\text{PMNS}{\sqrt{\tilde{m}_i}} O {\sqrt{\mu^{-1}}}\;,
\end{equation}
where $O$ is an arbitrary orthogonal $3\times 3$ matrix and can be parametrized as the product of three rotation matrices. In order to reduce the number of free parameters (degrees of freedom) of the model, one can make use of the ``minimal flavor violation hypothesis" \cite{Forero:2011pc}, where  it is assumed that the flavor is violated only in standard Dirac Yukawa couplings. Therefore,   $\mu$  is considered to be  diagonal.
\begin{figure}[H]
\begin{center}
\includegraphics[width=7.5cm,height=7cm]{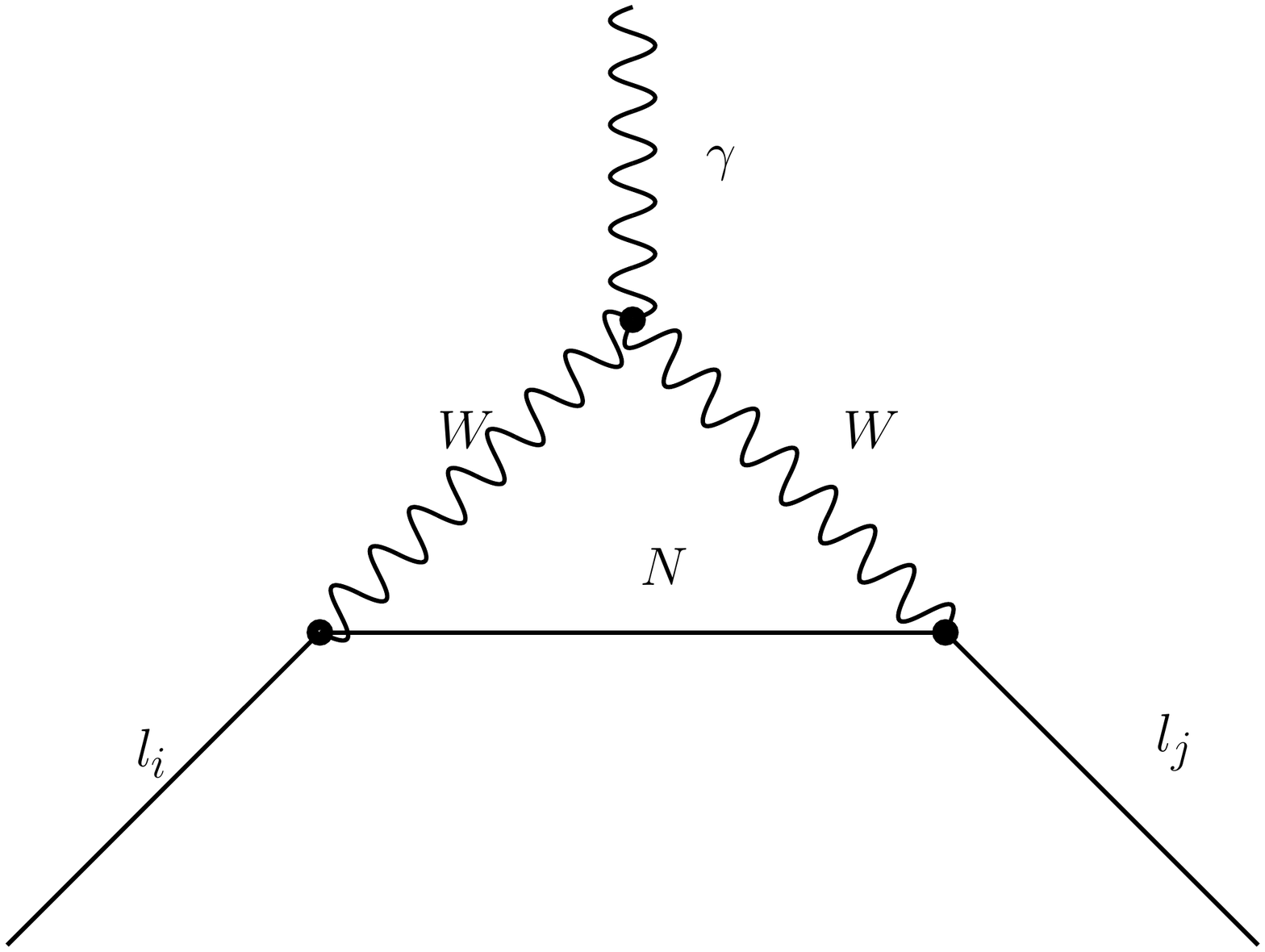}
\caption{Feynman diagram for $l_i \to l_j \gamma$ decay in seesaw model.}
\label{fd}
\end{center}
\end{figure}
Alternatively, one can also have inverse type-I seesaw mechanism by including three extra $SU(2)_L$ singets $S_i$, charged under $U(1)_L$ global symmetry as discussed in Ref. \cite{Forero:2011pc}. After electroweak symmetry breaking one obtains the mass matrix in the basis $(\nu, \nu^c,S)$ as   
\begin{equation}
\mathcal{M}_{\nu}=
\left( \begin{array}{ccc}
0&M_D&0\\
M_D^T&0&M\\
0&M^T&\mu\\
\end{array}\right),
\end{equation}
which gives the light neutrino mass as given in Eqn. (\ref{iss}).

Thus, in these low-scale inverse seesaw models, the  diagonalization of symmetric neutrino mass matrix ($\mathcal{M}_{\nu}$) leads to three light Majorana eigenstates $\nu_i$ with $i=1,2,3$ and six heavy neutrino states $s_j$ with $j=4,..,9$ (TeV scale). As a result, the active neutrino flavor state becomes,
\begin{equation}
\nu_{\alpha} = N_{\alpha i} \nu_i +\Theta_{\alpha j} s_j = V'_{\alpha k}N_{k}\;,
\end{equation}
where $V'=[N, \Theta]$ and $N_k =(\nu_i,s_j)^T$ with $k=1,2 \dots ,9$. Therefore, the  effective Lagrangian of  charged current weak interaction of neutrino mass eigenstate  is given by \cite{Schechter:8082pdf}
\begin{equation}
\mathcal{L}_{CC} \supset i \frac{g}{\sqrt{2}}\bar{l}_{\beta}K_{\beta\alpha}\gamma_{\mu}\nu_{\alpha L}W^{\mu}+{\rm h.c.},
\end{equation}
where 
\begin{equation}
K_{\beta\alpha} = \sum_{i=1}^9 \Omega^*_{i\beta}V'_{i\alpha}\;,
\end{equation}
with $\Omega$ as the $3 \times 3$ unitary matrix which diagonalizes the charged lepton mass matrix. Furthermore, if one considers the charged lepton mass matrix to be diagonal, then the $\Omega$ is simply an identity matrix and hence, $K=V'$.

In view of the fact that inverse seesaw model allows large light-heavy neutrino mixing, this gives rise to significant contributions to lepton flavor violating (LFV) decays ($l_i \to l_j \gamma$) via exchange of heavy neutrinos at one loop level \cite{LFV-loop1,LFV-loop2} as shown in the Fig \ref{fd}, and the corresponding one-loop contribution to the branching fraction for LFV decays is given by \cite{Ilakovac:1994kj}
\begin{equation}
\text{Br}(l_i \to l_j \gamma) = \frac{\alpha_W^3s_W^2}{256\pi^2}\frac{m_{l_i}^5}{M_W^4}\frac{1}{\Gamma_{l_i}}|G_{ij}^W|^2,
\end{equation}
where $G_{ij}^W$ is the loop function whose analytic form is
\begin{eqnarray}
G_{ij}^W &=& \sum_{k=1}^9 \Theta_{ik} \Theta_{jk}^* G_{\gamma}^W \left (\frac{m_{N_k}^2}{M_W^2}\right )~~~~~\text{with}  \nonumber\\
G_{\gamma}^W (x)&=& \frac{1}{12(1-x)^4}(10-43x+78x^2-49x^3 +4x^4)\;.
\end{eqnarray}
It should be noted that the non-unitarity parameters can be constrained by using the existing bound on the LFV decays. Such constraints on non-unitary parameters in low-scale seesaw mechanism (both inverse and linear seesaw mechanisms) are obtained in \cite{Forero:2011pc}, where the  mass of right handed neutrinos and the sterile fermions taken to be of the order of 1 TeV and 1 keV, and the obtained bounds on each parameter are summarized in the Table \ref{bounds}.
 \begin{table}[!htb]
  \centering
\begin{tabular}{|c| c c |c c| c c|} \hline
 Process &~~~~~~~~~ $\mu \to e \gamma$ & & ~~~~~~~~$\tau \to e \gamma$ & & ~~~~~~~~~$\tau \to \mu \gamma$ &\\ \hline
 Hierarchy & NH & IH & NH & IH & NH & IH	\\ \hline
$|\eta_{12}|<$  & $ 1.4\times 10^{-3}$ ~& $ 1.4\times 10^{-3}~$ & $2.8\times 10^{-2}$ & $2.8\times 10^{-2}$ ~& $2.8\times 10^{-2}$ & $2.8\times 10^{-2}$~\\
 $|\eta_{13}|<$  & $ 2.0\times 10^{-2}$ & $2.1\times 10^{-2}$ & $1.1\times 10^{-2}$ & $1.1\times 10^{-2}$ & $3.1\times 10^{-2}$ & $3.2\times 10^{-2}$\\
 $|\eta_{23}|<$  & $2.7\times 10^{-2}$ & $2.5\times 10^{-2}$ & $6.4\times 10^{-2}$ & $4.3\times 10^{-2}$ & $1.2\times 10^{-2}$ & $1.2\times 10^{-2}$\\ \hline  
\end{tabular}
\caption{\label{bounds}  Limits on unitarity 
  violation parameters from lepton flavor violation searches \cite{Forero:2011pc}.}
\end{table}
It can be seen from Table \ref{bounds} that the bounds on non-unitary lepton mixing parameters in the low scale inverse seesaw model is significantly large and thus, they can be probed at long-baseline experiments.  

%
\section{Neutrino oscillation with Non-unitarity effects }

In this section, we discuss how  the neutrino oscillation probability gets modified in presence of non-unitary lepton mixing.  The time evolution equation of neutrino mass eigenstates in standard paradigm is given by
\begin{equation}
i\frac{d}{dt}|\nu_i \rangle = \mathcal{H}_m |\nu_i \rangle\;,
\end{equation}
 where $\mathcal{H}_m$ is Hamiltonian in presence of matter effect, which is given by
 \begin{eqnarray}
 \mathcal{H}_m =\frac{1}{2E} \left( \begin{array}{ccc}
 0&0&0\\
 0&\Delta m_{21}^2&0\\
 0&0&\Delta m_{31}^2\\
 \end{array} \right)+ U_\text{PMNS}^{\dagger} \left( \begin{array}{ccc}
 V_\text{CC}+V_\text{NC}&0&0\\
 0&V_{NC}&0\\
 0&0&V_{NC}\\
 \end{array} \right) U_\text{PMNS}\;,
 \end{eqnarray}
with $\Delta m_{ij}^2= m_i^2-m_j^2$, $V_\text{CC}=\sqrt{2}G_F n_e$ and $V_\text{NC}=-G_F n_n/\sqrt{2}$ are the charged current and neutral current matter potentials respectively. In presence of non-unitary lepton mixing, the charged current and neutral current interaction Lagrangian gets modified as \cite{Antusch:2006vwa}
\begin{equation}
-\mathcal{L}_{int} =V_\text{CC} \sum_{i,j} N^*_{ei}N_{ej} \bar{\nu}_i \gamma^0 \nu_j +V_\text{NC}\sum_{\alpha,i,j}N^*_{\alpha i}N_{\alpha j}\bar{\nu}_i
\gamma^0\nu_j\;,
\end{equation} 
 which yields the effective Hamiltonian as
 \begin{eqnarray}
 \mathcal{H}^N_m =\frac{1}{2E} \left( \begin{array}{ccc}
 0&0&0\\
 0&\Delta m_{21}^2&0\\
 0&0&\Delta m_{31}^2\\
 \end{array} \right)+ N^{\dagger} \left( \begin{array}{ccc}
 V_\text{CC}+V_\text{NC}&0&0\\
 0&V_\text{NC}&0\\
 0&0&V_\text{NC}\\
 \end{array} \right) N.
 \end{eqnarray}
Then the oscillation probability after travelling a distance L can be obtained as
\begin{equation}
P_{\alpha \beta} (E,L) =|\langle\nu_{\beta}|\nu_{\alpha}(L)|^2 = \left |\left(N e^{-i{\cal H}_m^N L}N^{\dagger}\right)_{\beta \alpha}\right |^2.
\end{equation} 
The  non-unitarity  effects  originating  from  the  heavy and active  neutrino  mixing  can also  be
parametrized as
\begin{equation}
N=TU=(I-\alpha)U\;,\label{tm}
\end{equation}
where $U$ is the unitary matrix equivalent to standard neutrino mixing matrix and $T$ is lower triangular matrix. The  unitarity violating matrix can be of the form
\begin{equation}
T=\left( \begin{array}{ccc}
\alpha_{11}&0&0\\
\alpha_{21}&\alpha_{22}&0\\
\alpha_{31}&\alpha_{32}&\alpha_{33}\\
\end{array}\right).
\end{equation}
It should be noted from  Eqn. (\ref{tm}) that the diagonal elements of $T$ are of the form $ (1-\alpha_{ii}) \to \alpha_{ii} $.  %
Moreover, the relation between the parameters in two parametrizations of non-unitary mixing is obtained in \cite{Blennow:2016jkn} and it is given by
\begin{equation}
\left( \begin{array}{ccc}
\eta_{11}&0&0\\
2\eta_{12}^*&\eta_{22}&0\\
2\eta_{13}^*&2\eta_{23}^*&\eta_{33}\\
\end{array}\right)= 
\left( \begin{array}{ccc}
\alpha_{11}&0&0\\
\alpha_{21}&\alpha_{22}&0\\
\alpha_{31}&\alpha_{32}&\alpha_{33}\\
\end{array}\right).
\end{equation}
As the triangular parametrization is the preferred one for oscillation studies, we use these relations while doing the analysis. We use the General Long Baseline Experiment Simulator (GLoBES)~\cite{Huber:2004-1,Huber:2009-2} package along with the plugin MonteCUBES \cite{Blennow:2009} in order to do the numerical calculations. The  neutrino oscillation parameters which we use in our analysis are given in the Table \ref{osc1}. Further, we use the non-unitarity parameters which satisfy the constraints that are given in Table \ref{bounds} and the values that we use in the analysis are given in Table \ref{nup}.
\begin{table}
\begin{center}
\begin{tabular}{|c|c|c|} \hline
Parameters & Best fit & 3$\sigma$ range\\ \hline 
 $\sin^2\theta_{12}$ & 0.321 &\\
   $\sin^2 2\theta_{13}$ & 0.084 & \\
   $\sin^2 \theta_{23}$ (LO) & 0.44 & ~[0.38:0.50]~ \\
   $\sin^2 \theta_{23}$ (HO) & 0.56 & ~[0.50:0.62]~\\
   $\Delta m_{atm}^2$ (NH)& $2.5 \times 10^{-3} ~{\rm eV}^2$&  [2.38:2.62]$\times 10^{-3} ~{\rm eV}^2$\\
   $\Delta m_{atm}^2$ (IH) & $-2.5 \times 10^{-3} ~{\rm eV}^2$&~ $[-2.62:-2.38]\times 10^{-3} ~{\rm eV}^2$ \\
   $\Delta m_{21}^2$&  $7.56 \times 10^{-5}~ {\rm eV}^2$&  \\
   $\delta_{CP} $ & $-90^\circ$& ~$[-180^\circ:180^\circ]$~\\
\hline
\end{tabular}
 \caption{{\label{osc1}}The values of neutrino oscillation parameters used in the analysis \cite{Valle:2017pdf}.}
\end{center}
\end{table}
\begin{table}[!htb]
  \centering
\begin{tabular}{|c| c | c|  c|} \hline 
 Process & $\mu \to e \gamma$ & $\tau \to e \gamma$ & $\tau \to \mu \gamma$\\ \hline
$|\eta_{12}|$  &~ $ 1.2\times 10^{-3}$ ~& ~$2.6\times 10^{-2}$ ~ &~ $2.6\times 10^{-2}$~\\
 $|\eta_{13}|$  &~ $ 1.8\times 10^{-2}$  ~& ~$1\times 10^{-2}$~ & ~ $3\times 10^{-2}$~\\
 $|\eta_{23}|$  &~ $2.5\times 10^{-2}$~ & ~$4.2\times 10^{-2}$ ~&~  $1\times 10^{-2}$~\\
  \hline 
\end{tabular}
 \caption{{\label{nup}}The values of non-unitarity parameters  used in the analysis.}
\end{table}
The phases associated with the complex non-unitarity parameters can vary from $-\pi$ to $\pi$. However, we assume these phases to be zero while doing the analysis unless otherwise mentioned.

As the long-baseline experiments are mainly looking for $\nu_{\mu} \to \nu_{e}$  and $\bar{\nu}_{\mu} \to \bar{\nu}_{e}$  oscillations, first of all, we would like to see  relative deviation in the $\nu_{\mu} \to \nu_{e}$ oscillation probability due to  the  unitarity violation in lepton mixing. In order to do  this, we define a quantity $\Delta P_{\mu e} = \displaystyle{\frac{|P_{\mu e}^{NU}-P_{\mu e}^{SO}|}{P_{\mu e}^{SO}}}$, where $P_{\mu e}^{NU}$ is the oscillation probability with unitarity violation and $P_{\mu e}^{SO}$ is the oscillation probability in standard three  flavor oscillation framework. We obtain the quantity $\Delta P_{\mu e}$ for different energy and baseline. While doing the numerical calculation, we assume that the atmospheric mixing angle is maximal ($\sin^2\theta_{23}=0.5$) and use the values of non-unitarity parameters as given in Table III.
\begin{figure}
\begin{center}
\minipage{0.25\textwidth}
    \includegraphics[height=5cm,width=7cm]{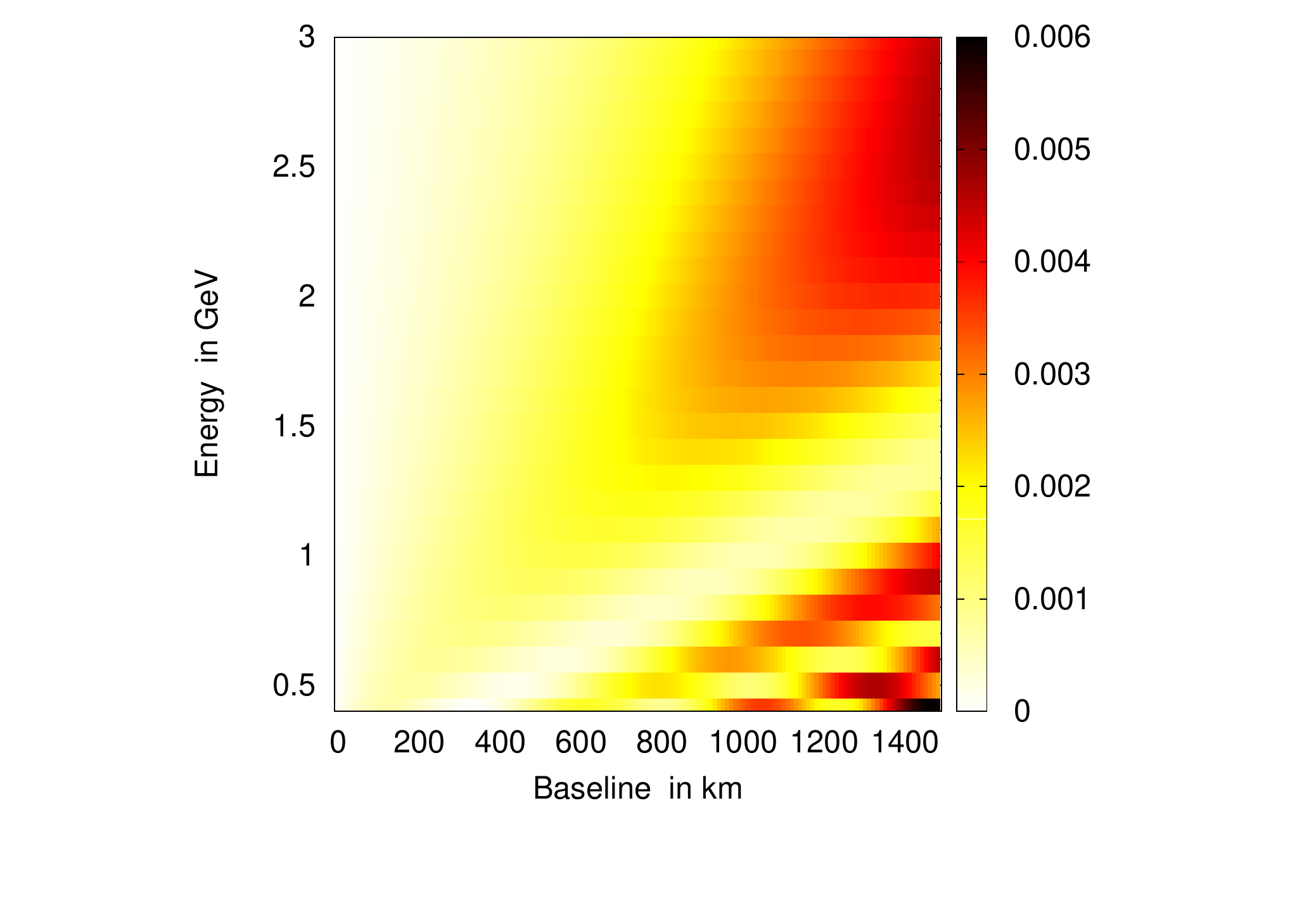}\\
    \includegraphics[height=5cm,width=7cm]{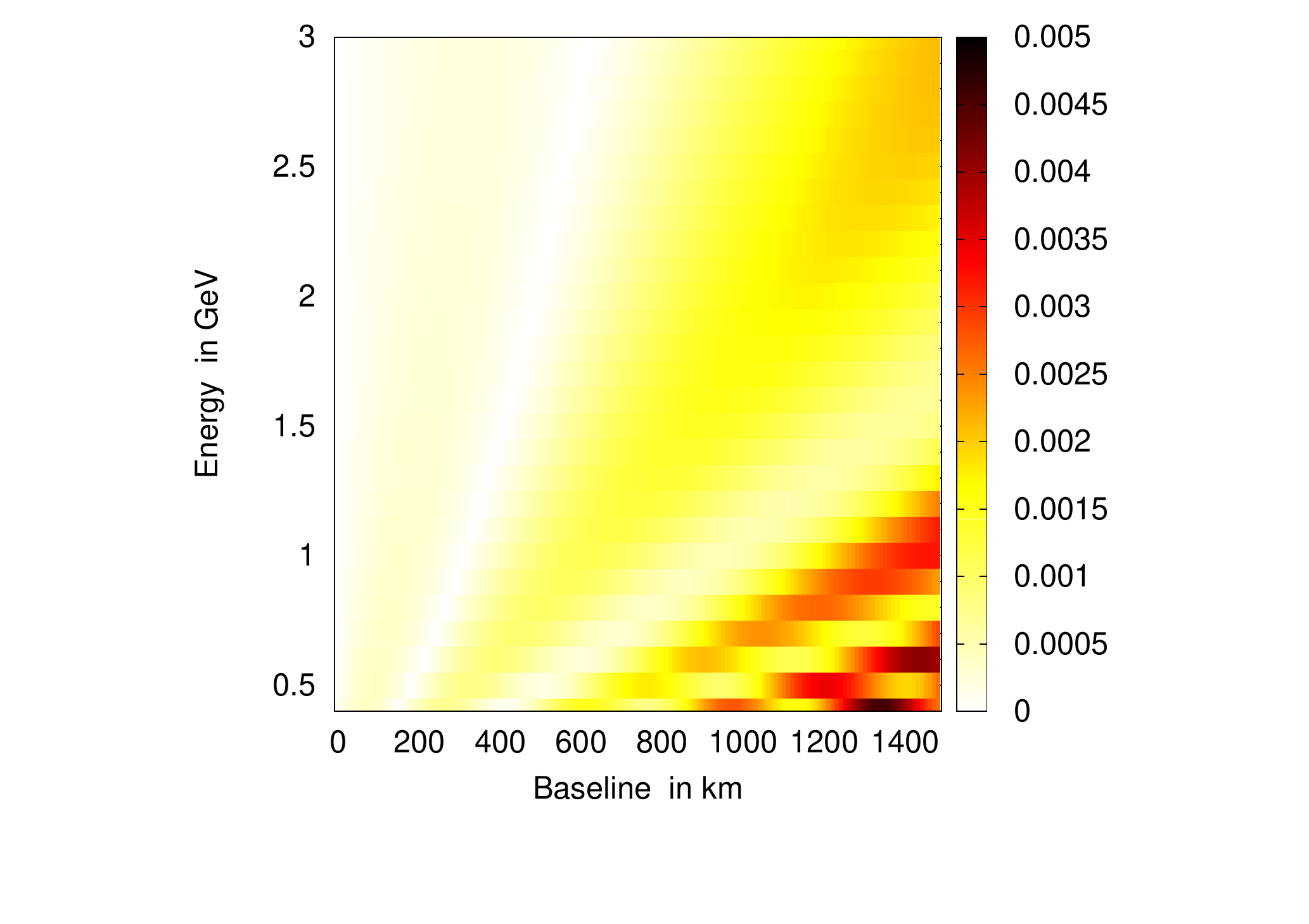}
    \endminipage\hfill
\minipage{0.25\textwidth}
    \includegraphics[height=5cm,width=7cm]{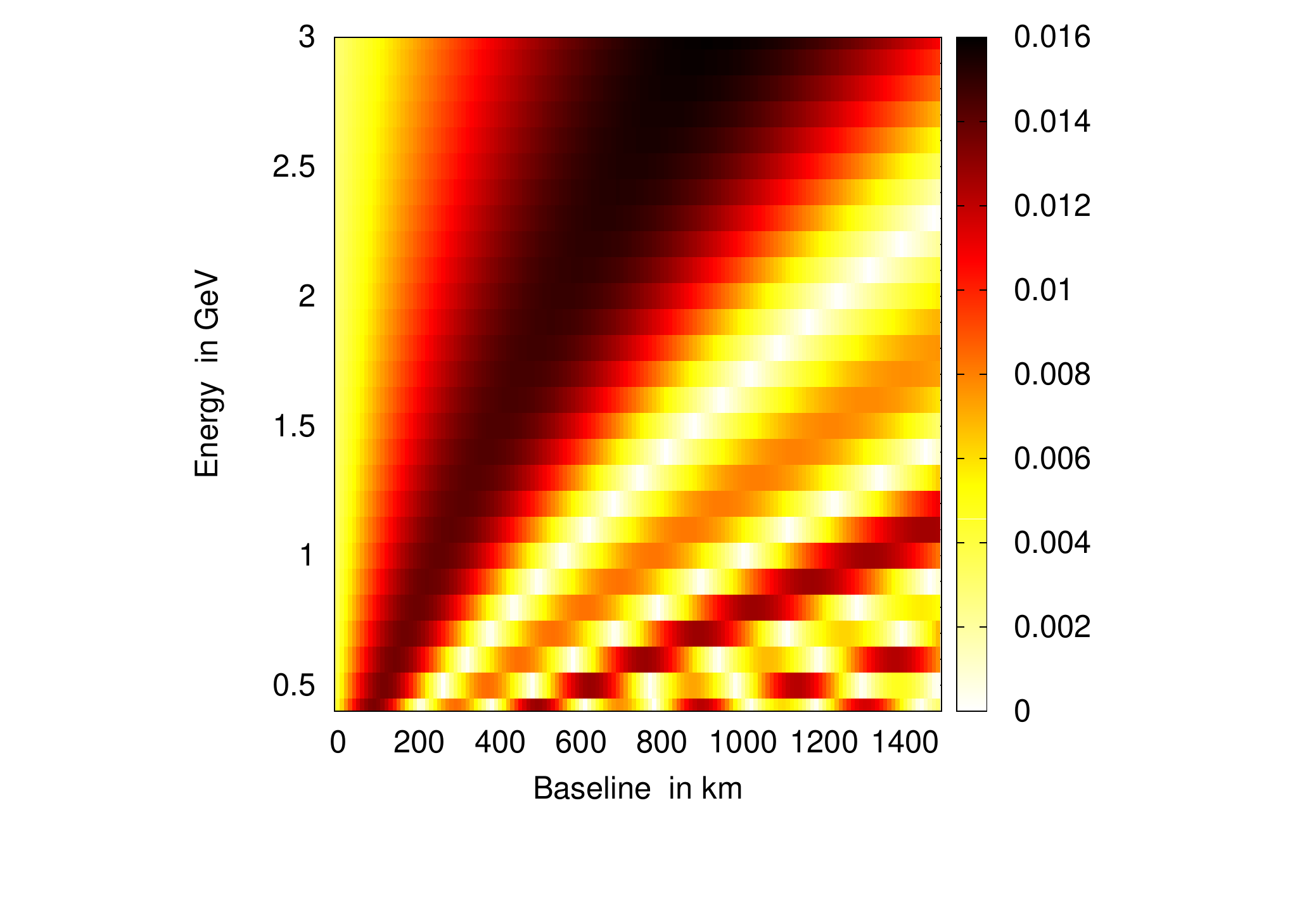}\\
    \includegraphics[height=5cm,width=7cm]{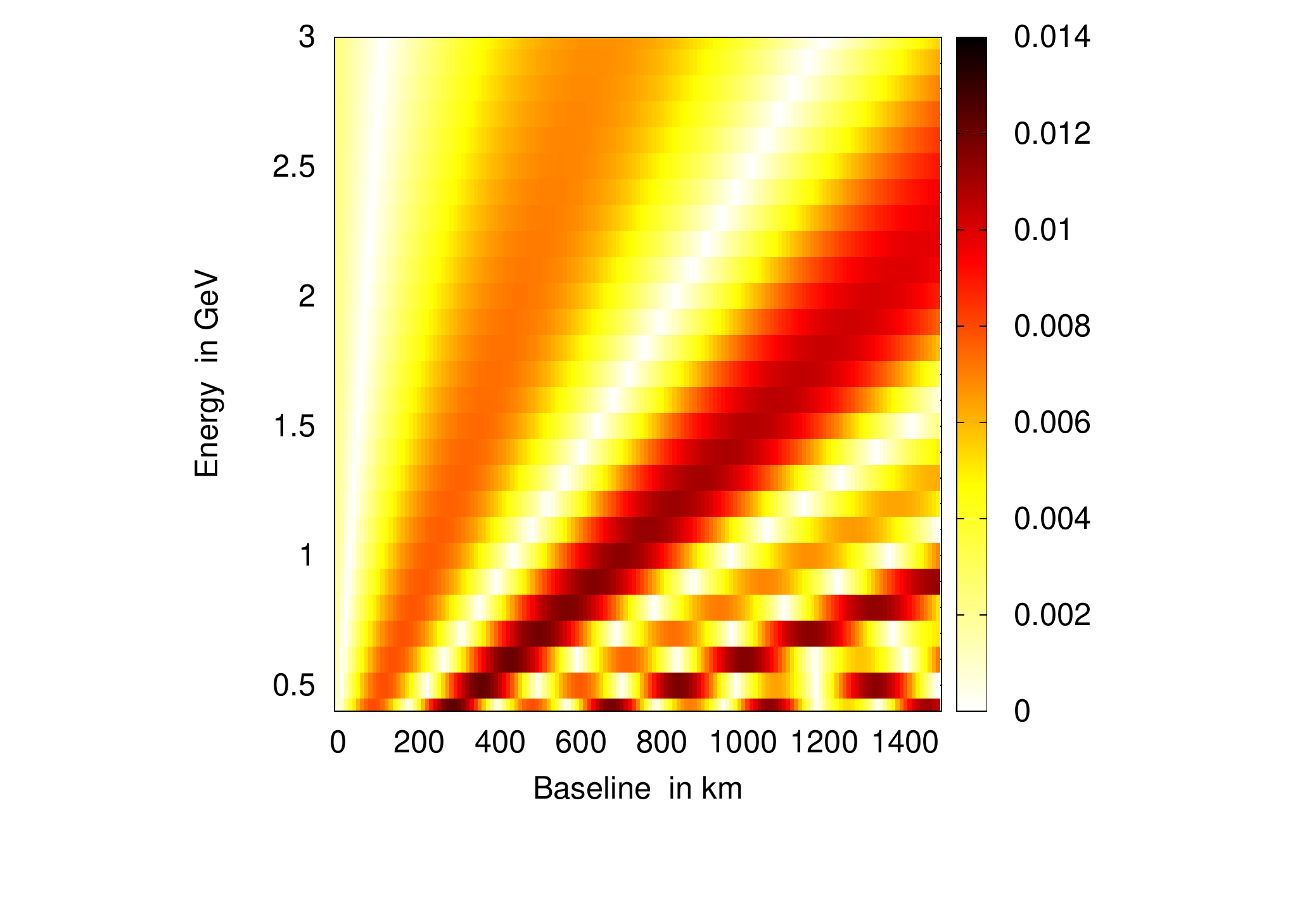}
    \endminipage\hfill
\minipage{0.35\textwidth}
    \includegraphics[height=5cm,width=7cm]{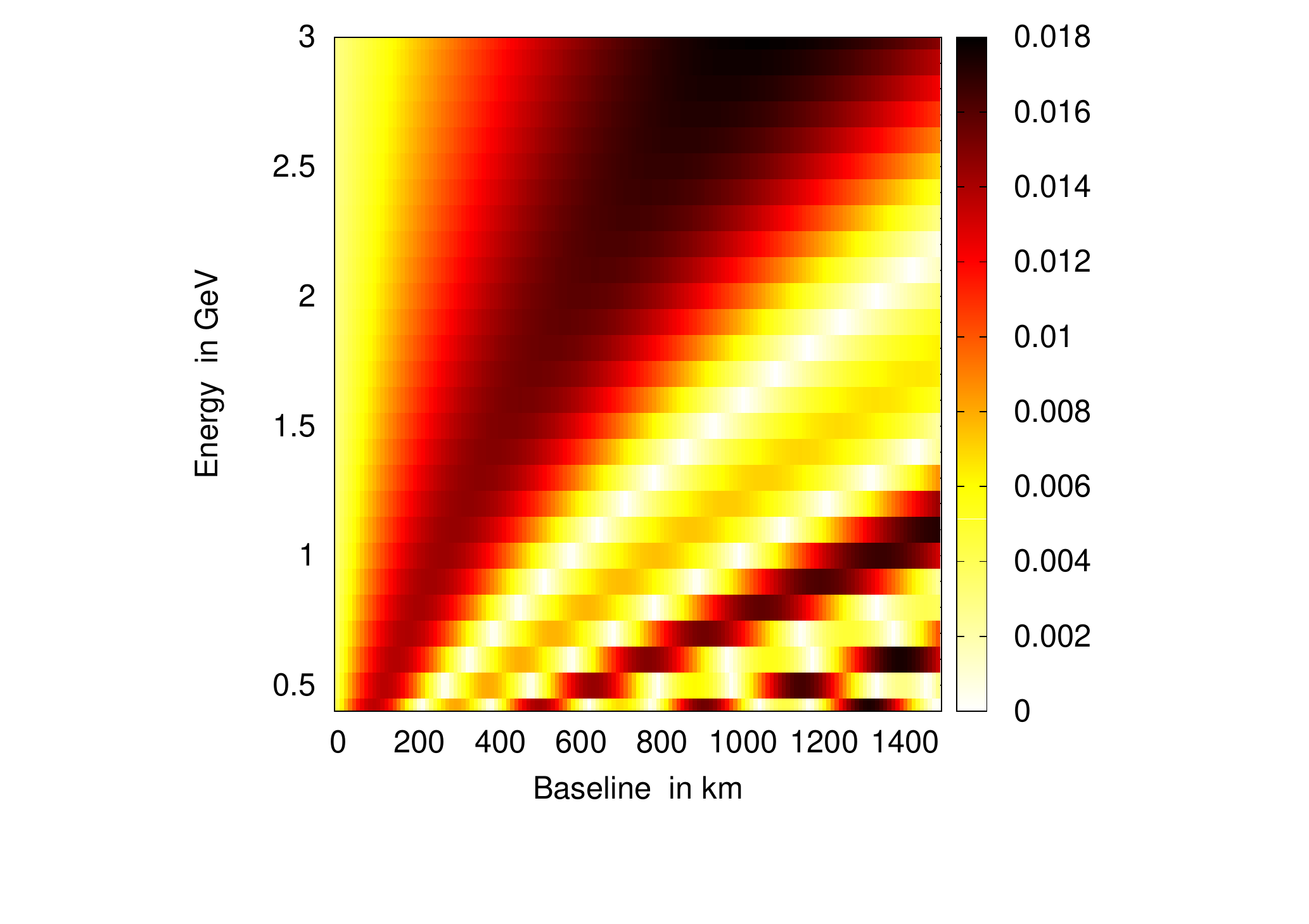}
    \includegraphics[height=5cm,width=7cm]{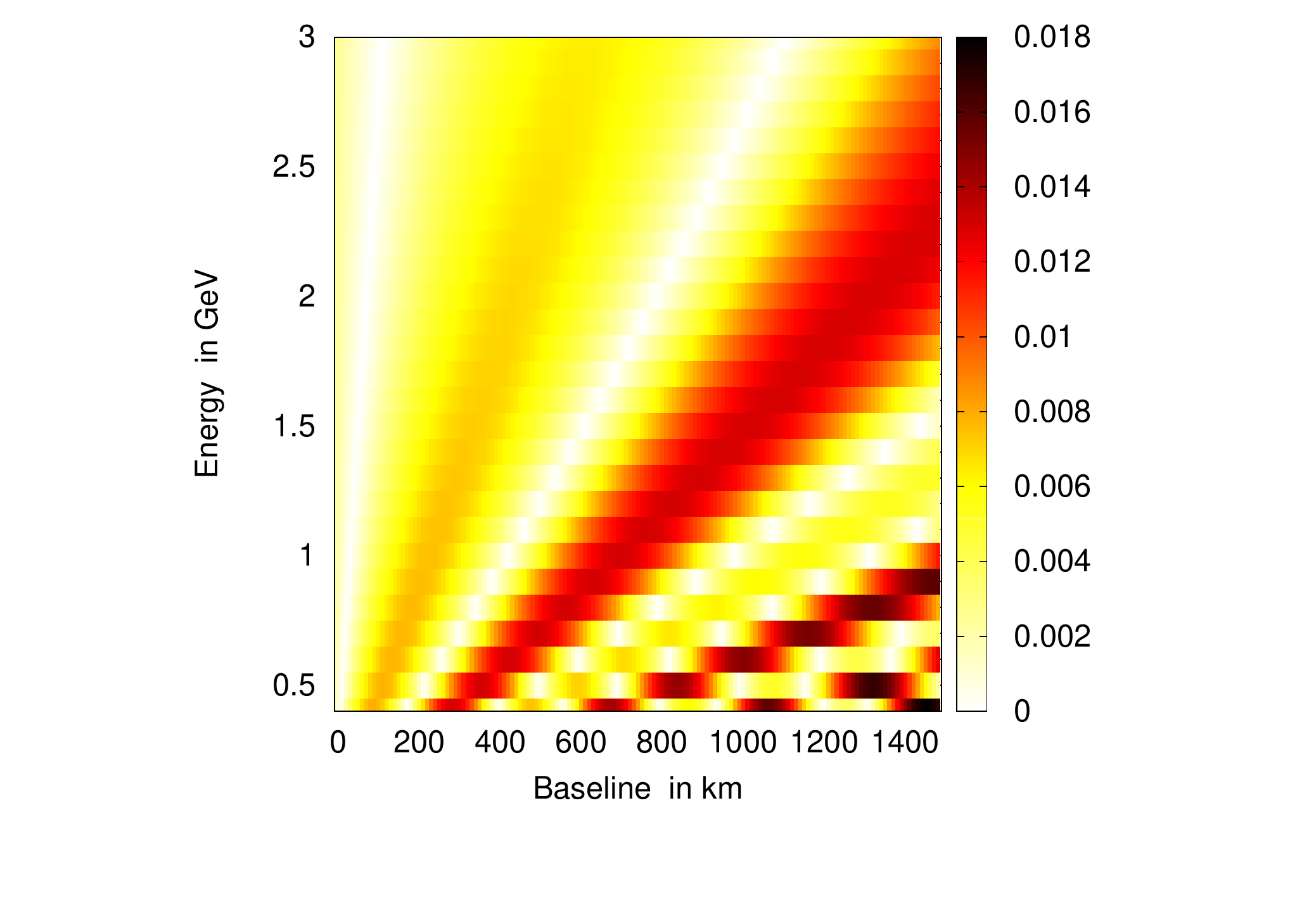}
    \endminipage
  \end{center}
\caption{{\label{le}} $\Delta P_{\mu e}$ in $L-E$ plane for the non-unitarity parameters which are constrained by $\mu \to e \gamma$ (left panel), $\tau \to e \gamma$ (middle panel) and $\tau \to \mu \gamma$ (right panel). In the top (bottom) panel the hierarchy of neutrino is assumed to be normal (inverted).}
\end{figure} 
 Fig. \ref{le} shows the variation of  $\Delta P_{\mu e}$ in $L-E$ plane for the non-unitarity parameters which are constrained by $\mu \to e \gamma$ (left panel), $\tau \to e \gamma$ (middle panel) and $\tau \to \mu \gamma$ (right panel). The darker regions  correspond to large amount of relative deviation in oscillation probability.  The bound on non-unitarity parameter
 $\eta_{12}$ (which plays the major role)  from $\mu \to e \gamma$  decay process is too constrained  and hence, the  parameters constrained by $\mu \to e \gamma$ do not have any significant contributions to $\Delta P_{\mu e}$ as  seen from the figure. Whereas, the bounds on the non-unitarity parameters
 (mainly $\eta_{12}$) are less constrained by other lepton flavor violating decay processes and they significantly contribute to $\Delta P_{\mu e}$.  Therefore, such non-unitarity parameters can be probed at long-baseline experiments like T2K (peak energy= 0.6 GeV, baseline = 295 km), NO$\nu$A (peak energy= 1.2 GeV, baseline = 810 km) and DUNE (peak energy= 2.5 GeV, baseline = 1300 km).  Moreover, these non-unitarity parameters play crucial role in the determination of oscillation parameters by these experiments. For simplicity, hereafter we focus on the non-unitarity parameters which are constrained by $\tau \to \mu \gamma$ process.

\section{Implications of non-unitary lepton mixing at LBL experiments}
Over the past few decades the knowledge of neutrino oscillation parameters 
$( \theta_{12}, \theta_{13}, \theta_{23}, \Delta m_{21}^2, \Delta m_{31}^2, \delta_{CP})$ within the standard three-flavor framework has improved dramatically. Howbeit, the leptonic CP phase, the mass hierarchy of neutrino (Normal: $\Delta m_{31}^2>0$  or Inverted: $\Delta m_{31}^2<0$) and the octant of atmospheric mixing angle (Lower Octant: $\theta_{23}<45^{\circ}$ or Higher Octant: $\theta_{23}>45^{\circ} $) are still  not known. The current status of neutrino oscillation parameters by including the latest results from T2K and NO$\nu$A experiments can be seen in \cite{Valle:2017pdf}. These recent experimental results hint towards $\delta_\text{CP}=-90^\circ$ and  show a slight preference for normal neutrino mass ordering, with $\Delta \chi^2 =2.7$. 
Moreover, the maximal mixing
of atmospheric mixing angle is disfavoured at $\Delta \chi^2 =6.0$ and the Lower Octant is preferred with $\Delta \chi^2 =2.1$
for normal neutrino mass ordering, whereas for inverted mass ordering the local minimum is in the Higher Octant with $\Delta \chi^2 =2.7$.
The current and future generation long-baseline experiments play crucial role in the resolution of these degeneracies among the oscillation parameters, which will eventually provide  a complete understanding of physics behind lepton mixing.  In this section, we mainly discuss  how the  non-unitary lepton mixing affect the determination of current unknowns in neutrino oscillation sector by considering NO$\nu$A as a case of study. 

NO$\nu$A uses an upgraded NuMI beam power of 0.7 MW at Fermilab. The Main Injector accelerator produces  mesons  by colliding  120 GeV proton beam on graphite target, which ultimately produce the neutrino beam through their decay. The produced neutrino beam is directed towards  14 kton totally active scintillator  detector (TASD) placed about 810 km away from Fermilab (near the Ash River). It also has a 0.3 kton near detector located at the Fermilab site to monitor the un-oscillated neutrino or anti-neutrino flux. Moreover, NO$\nu$A  makes use of off-axis technique to get neutrino energy spectrum with very narrow band. Therefore, the far detector of NO$\nu$A  experiment is placed $0.8^\circ$ off-axis from the NuMI beam line. In the analysis, we consider 3 years run each in neutrino and anti-neutrino modes which corresponds to a total of $6\times 10^{20}$ protons on target per year. The other experimental specifications of NO$\nu$A are taken from \cite{ska12} with the following characteristics:
\begin{itemize}
\item Signal efficiencies: 45\% for electron neutrino and electron anti-neutrino signals, whereas 100\% for both muon neutrino and muon anti-neutrino signals.
\item Background efficiencies: There are mainly three backgrounds and they are
 \begin{enumerate}
  \item Mis-ID muons acceptance:  The mis-identified muons (anti-muons) at the detector are about 0.83\% (0.22\%).
  \item NC background acceptance: There exist almost 2\% (3\%) neutral current events at the detector, which resemble the muon neutrino (muon anti-neutrino) events. 
  \item Intrinsic beam contamination: The possibility of existence of  electron neutrino (electron anti-neutrino) in the neutrino beam is about 26\% (18\%).
\end{enumerate}  
\end{itemize}
And we also assume that there exists $5 \%$ normalization error on signal and $10 \%$ on background. The migration matrices for NC background smearing are taken from \cite{ska12}. 

The following subsections discuss the discovery reach of non-unitarity parameters and their impacts on the determination of mass hierarchy, octant of atmospheric mixing angle and the CP-violating phase. At the end of this section, we also discuss about how the effect of non-unitarity mixing on the parameter degeneracy resolution capability of NO$\nu$A. 
\begin{figure}
 \begin{center}
  \includegraphics[height=5cm,width=7cm]{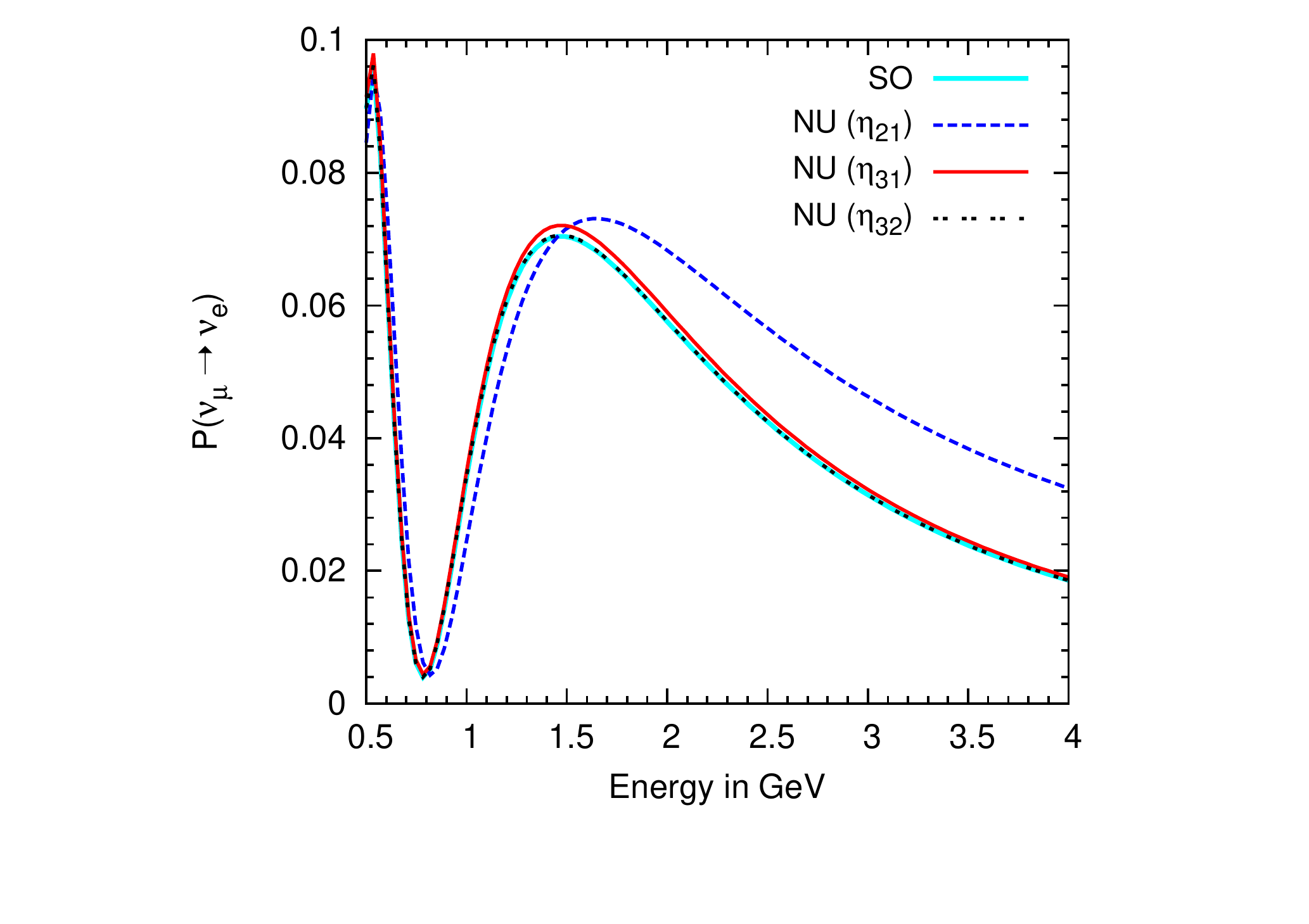}
 \includegraphics[height=5cm,width=7cm]{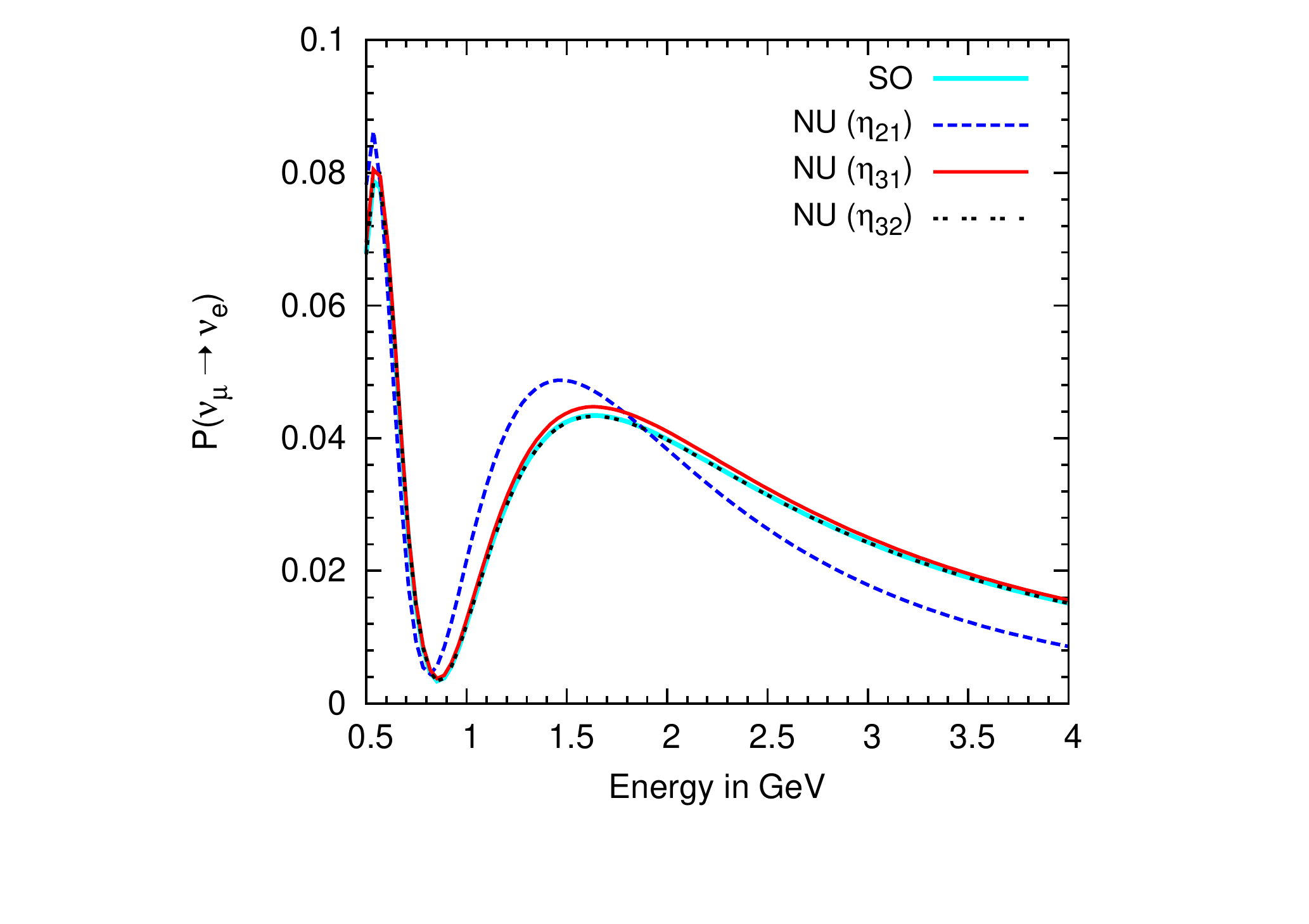}\\
 \includegraphics[height=5cm,width=7cm]{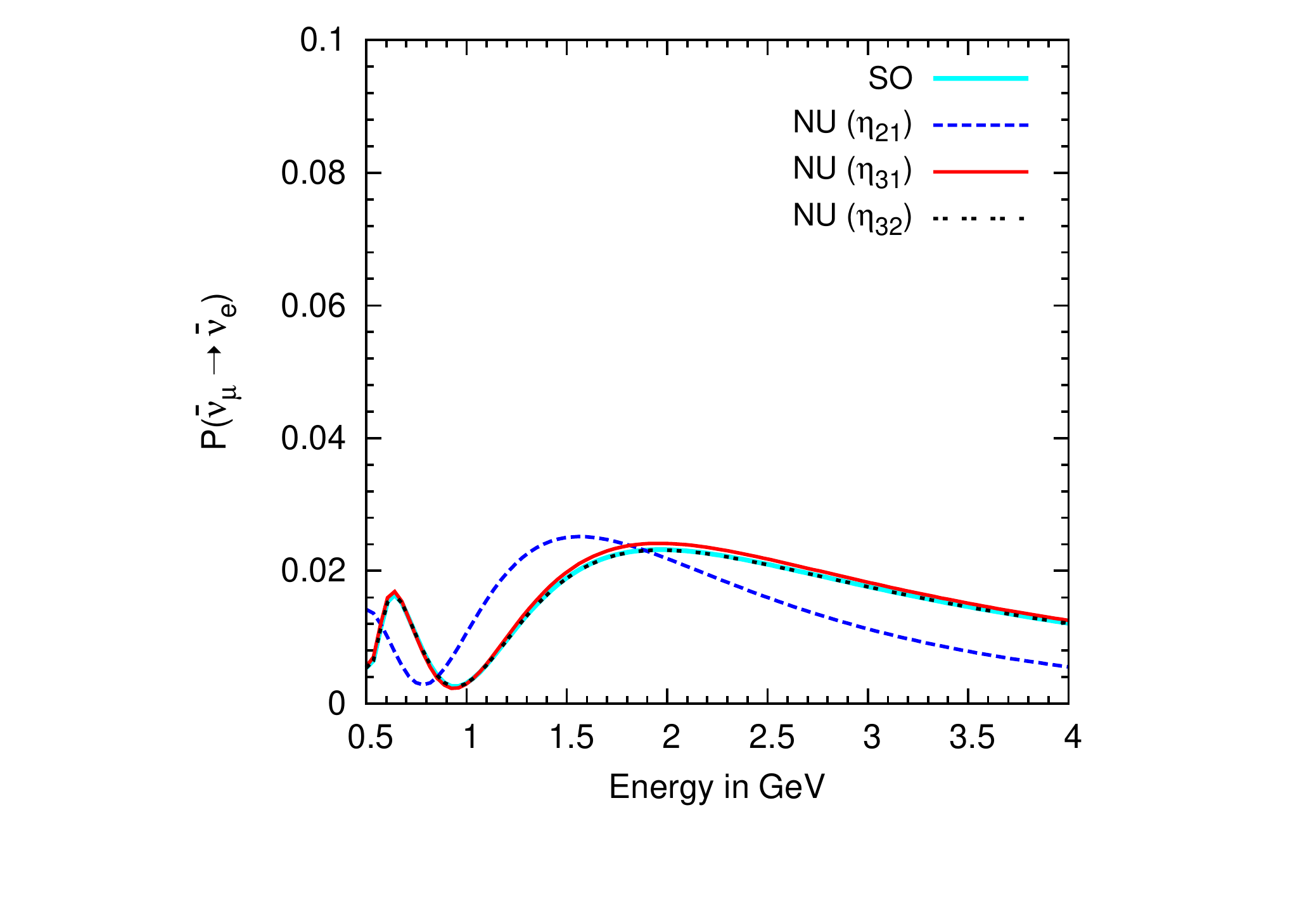}
 \includegraphics[height=5cm,width=7cm]{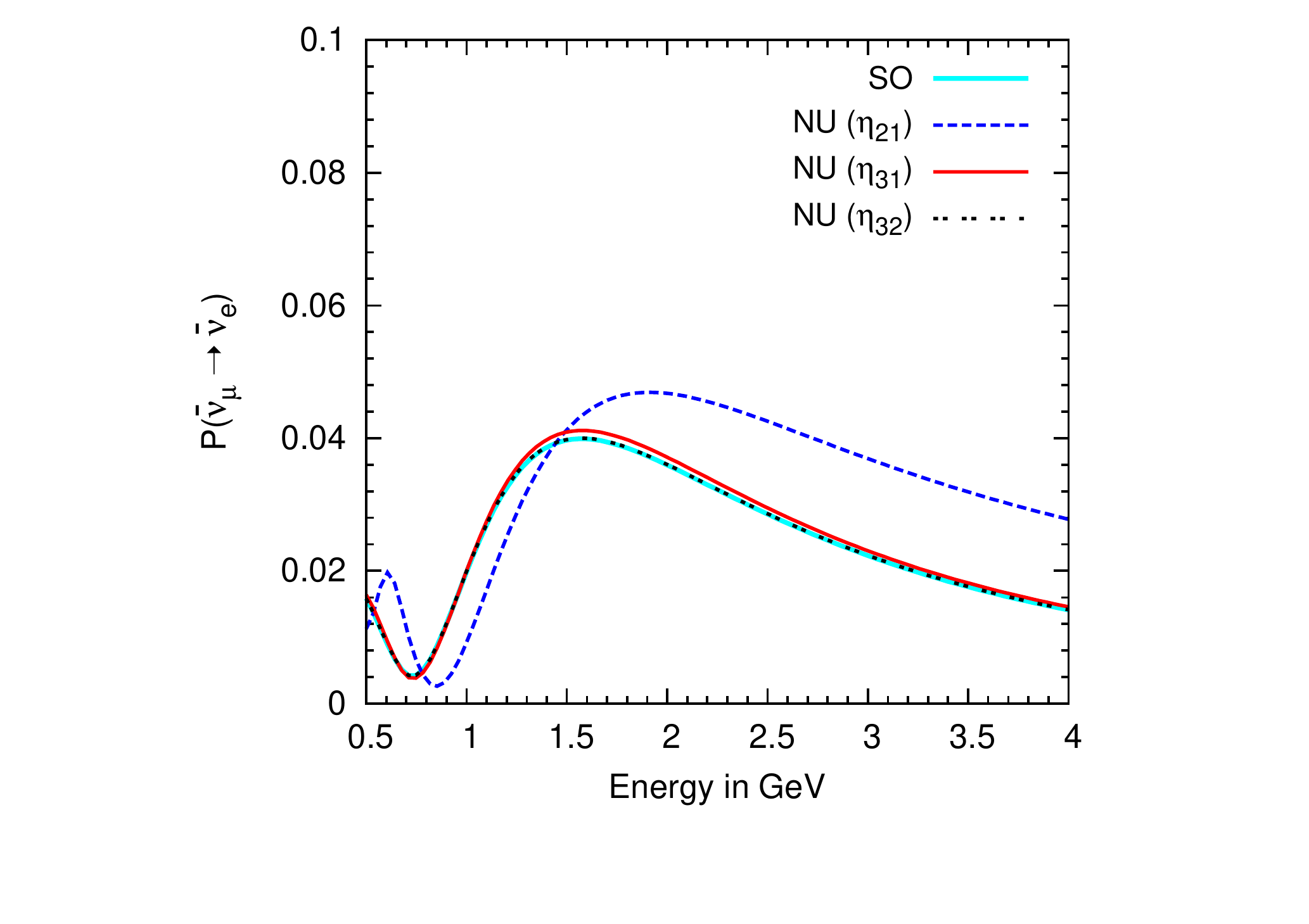}
 \caption{{\label{nuosc}}The neutrino (anti-neutrino) oscillation probability as a function of energy is given in the top (bottom) panel. The left (right) panel  corresponds to normal (inverted) hierarchy.}
 \end{center}
 \end{figure}
\subsection{Discovery reach of non-unitarity parameters}

As we are focusing on the non-unitarity parameters $\eta_{21},\eta_{31}, \eta_{32}$ and their corresponding CP-violating phases, it is most important to check how these parameters affect the oscillation probability. In the top (bottom) panel of Fig. \ref{nuosc}, we show the neutrino (anti-neutrino) oscillation probabilities as a function of neutrino energy. The left (right) panel  of the figure corresponds to oscillation probability for normal (inverted) hierarchy. We can see from the figure that the parameters $\eta_{31}$ and $\eta_{32}$ do not modify the oscillation probability significantly, whereas $\eta_{21}$ significantly modifies the oscillation probability. Therefore, the non-unitarity parameter $\eta_{21}$ can be probed at LBL experiments. However, one has to also take care of the role of phases associated with each non-unitarity parameters.
 
 \begin{figure}
\begin{center}
\minipage{0.25\textwidth}
    \includegraphics[height=5cm,width=7cm]{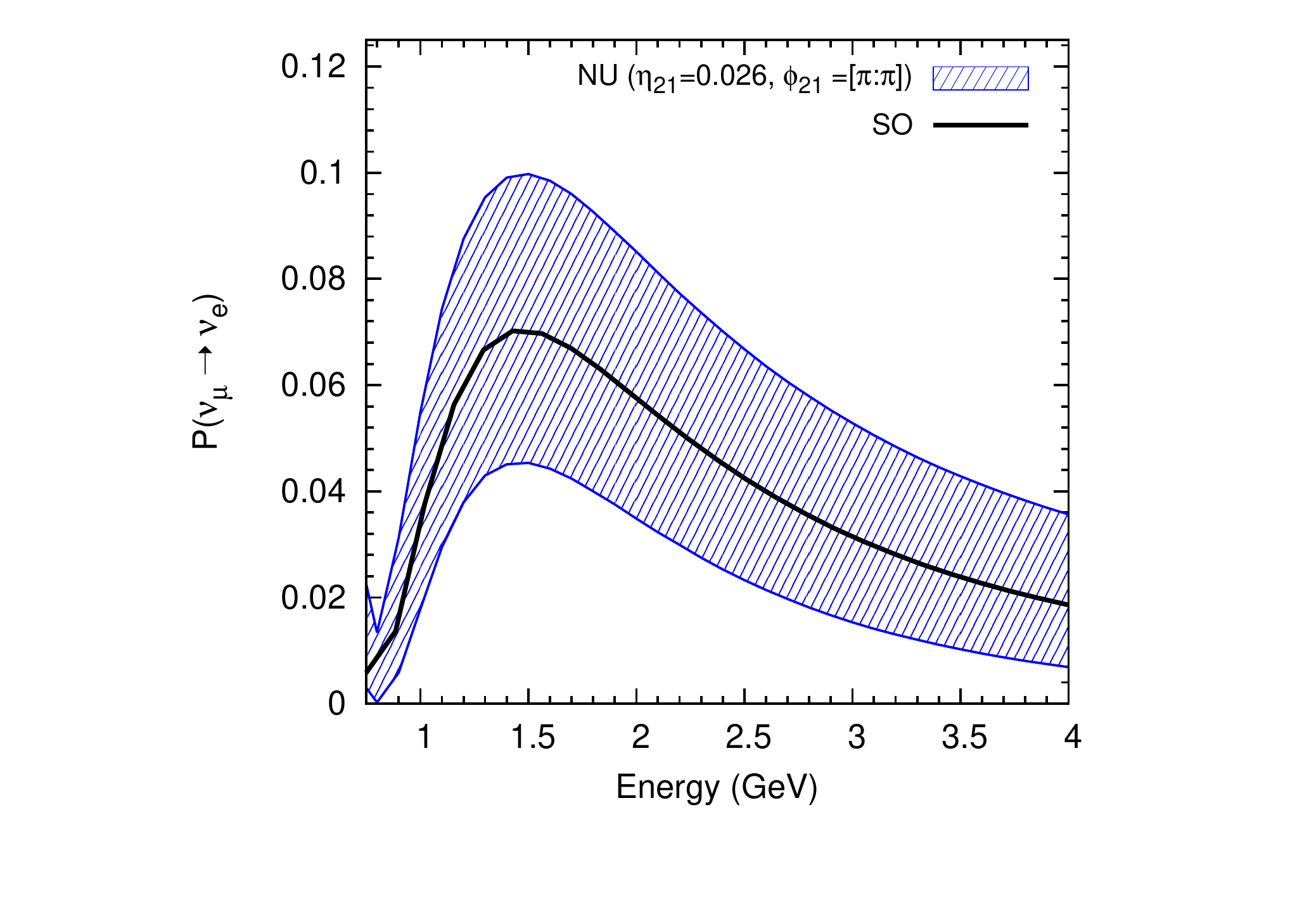}    \endminipage\hfill
\minipage{0.25\textwidth}
    \includegraphics[height=5cm,width=7cm]{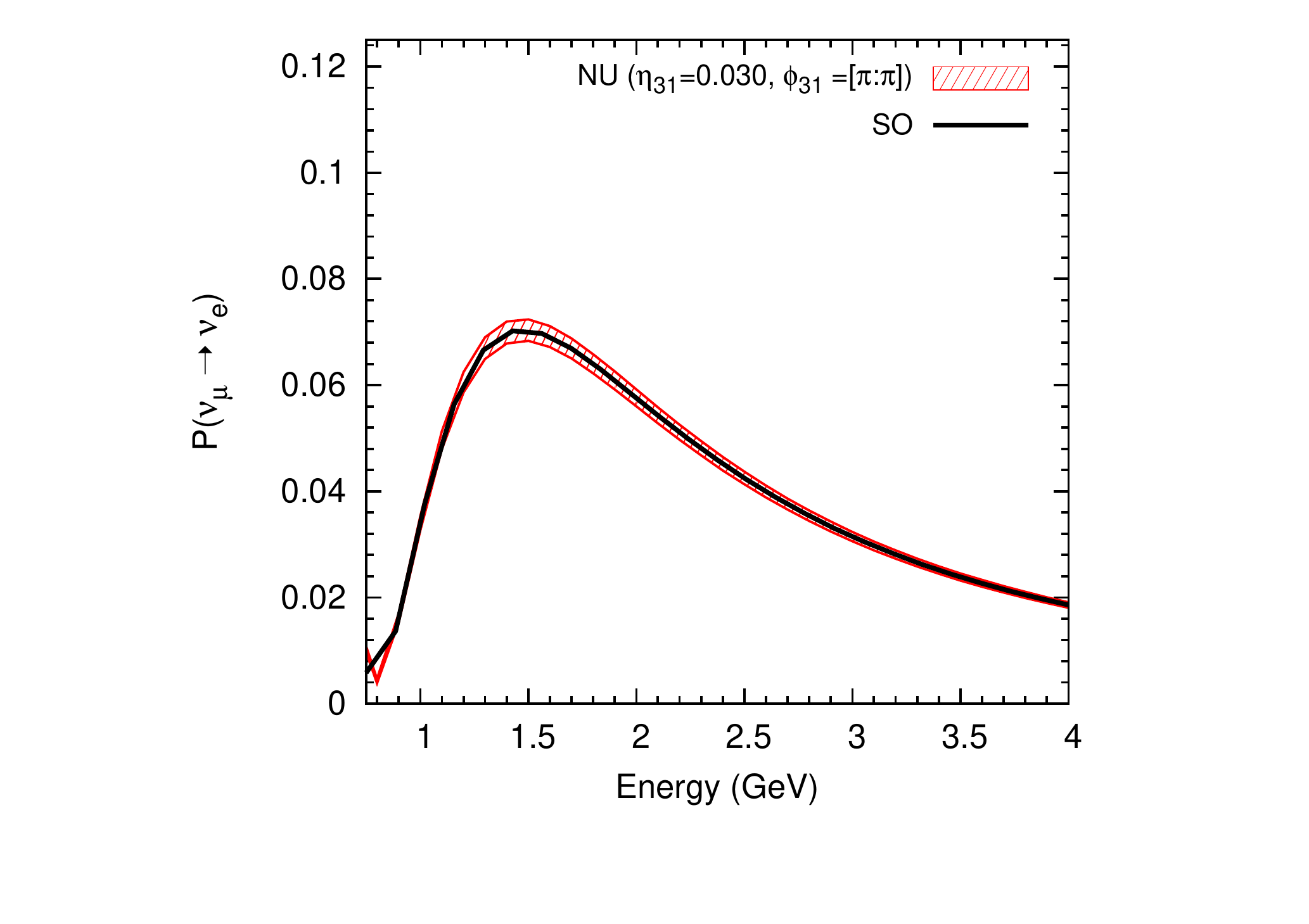}    \endminipage\hfill
\minipage{0.35\textwidth}
    \includegraphics[height=5cm,width=7cm]{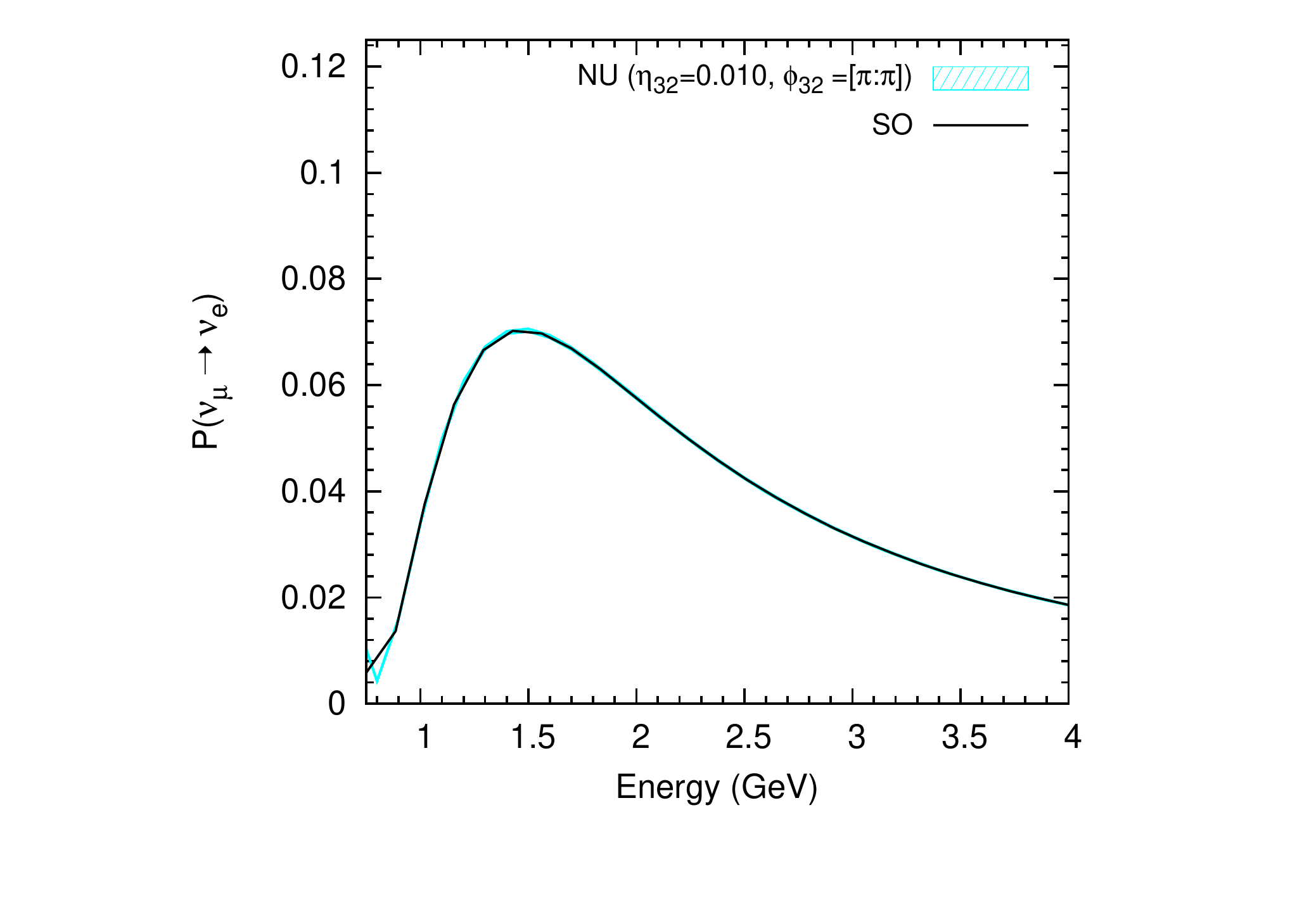}
    \endminipage
  \end{center}
\caption{{\label{nuphase}} The black curve corresponds to the oscillation probability in standard paradigm, whereas the blue, red, and cyan bands  correspond to oscillation probabilities in the presence of non-unitary parameters in 21, 31, and 32  sectors, respectively.}
\end{figure} 

 Fig. \ref{nuphase}  shows the oscillation probability in presence of CP-violating phases of non-unitarity parameters. In the figure, the black curve corresponds to the oscillation probability in standard paradigm, whereas the blue, red, and cyan bands correspond to oscillation probabilities in presence of non-unitary parameters in 21, 31, and 32  sectors, respectively. It can be seen from the figure that the non-unitarity parameter $\phi_{21}$  significantly alters the
oscillation probability, whereas the parameters $\phi_{31}$  and $\phi_{32}$ do not modify the oscillation
probability and their effect on oscillation probability is negligibly small. Thereby, it can be
easily understood from these figures that non-unitarity parameters in 21 sector, i.e, $\eta_{21}$ and $\phi_{21}$, play major role in the oscillation physics at long-baseline experiments. 
 
\begin{figure}
\begin{center}
\minipage{0.25\textwidth}
    \includegraphics[height=5cm,width=7cm]{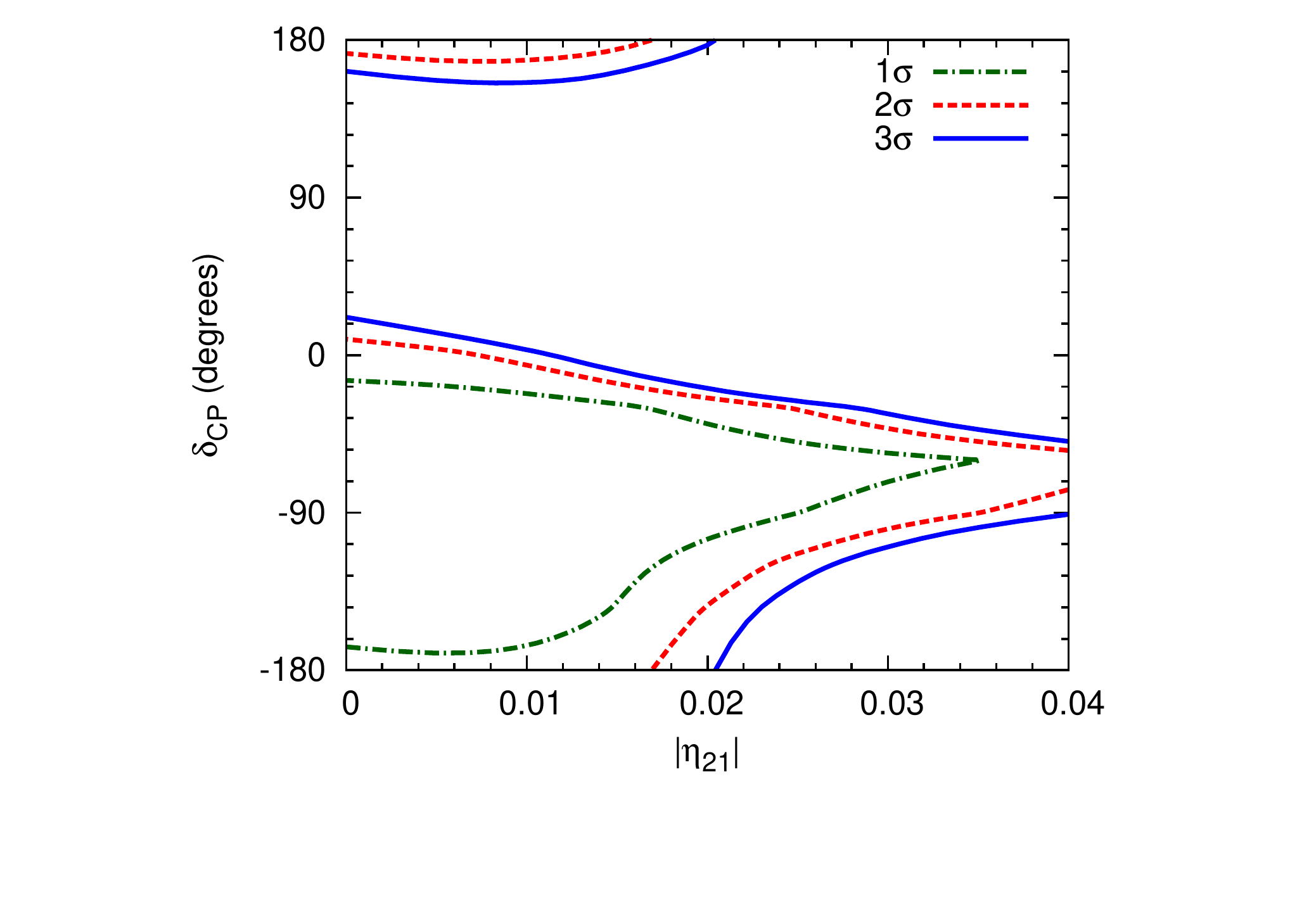}    \endminipage\hfill
\minipage{0.25\textwidth}
    \includegraphics[height=5cm,width=7cm]{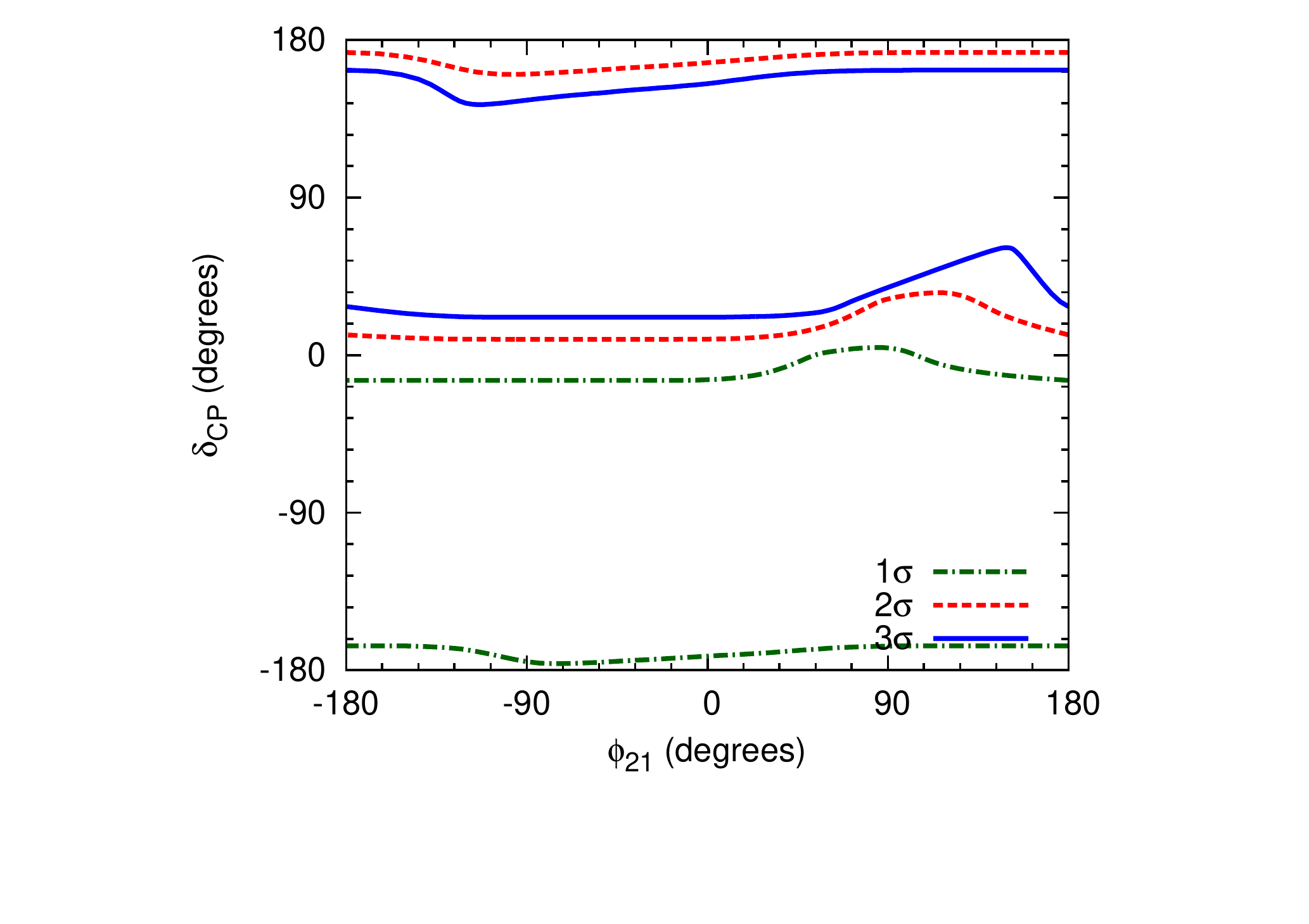}    \endminipage\hfill
\minipage{0.35\textwidth}
    \includegraphics[height=5cm,width=7cm]{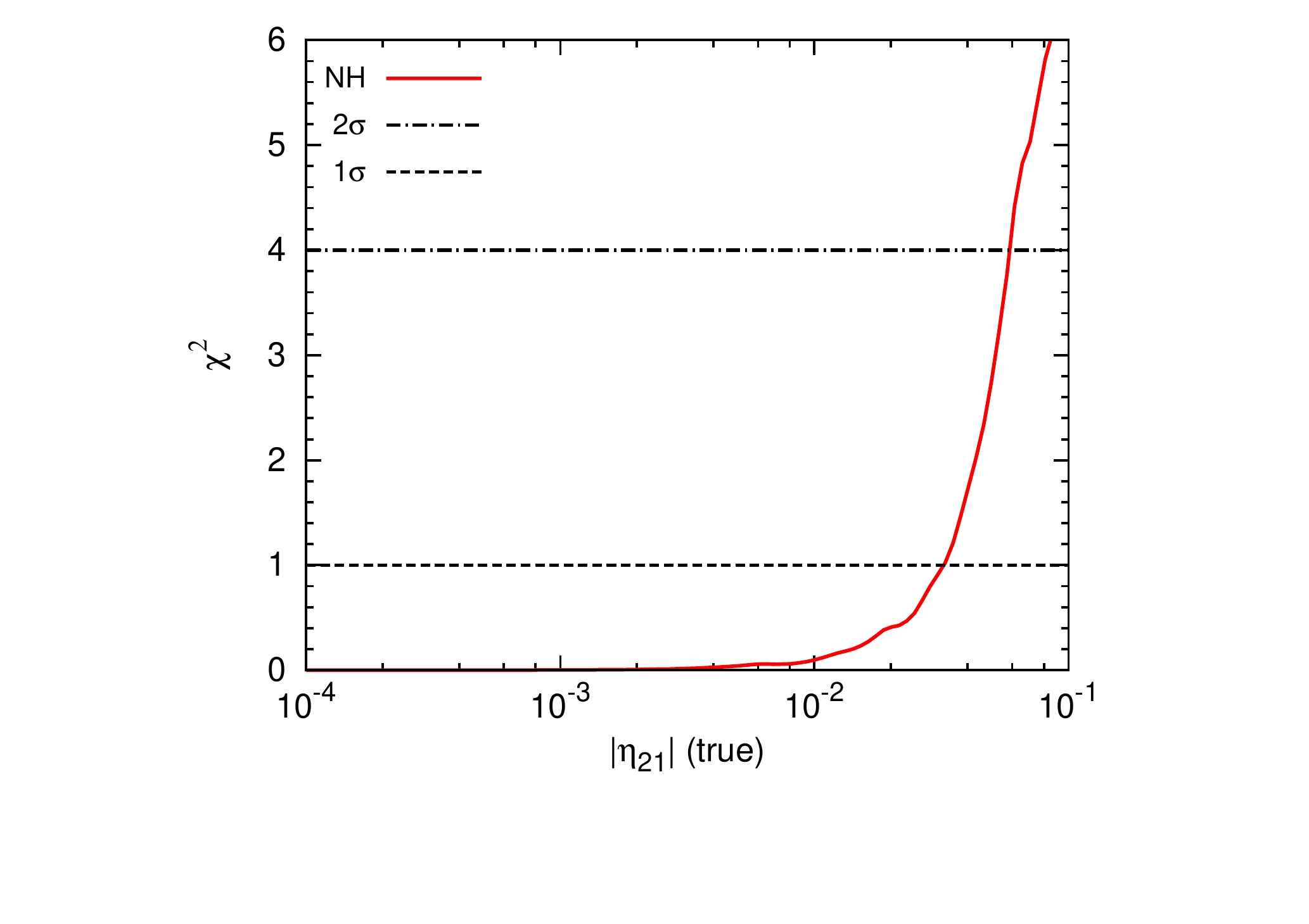}
    \endminipage
  \end{center}
\caption{{\label{allowedn}} The allowed parameter space in $|\eta_{21}|-\delta_{CP}$ ($\phi_{21}-\delta_{CP}$)  plane  is shown in left (middle) panel. The discovery reach of $|\eta_{21}|$ is shown in the right panel.}
\end{figure}

 Next, we analyse the potential of NO$\nu$A experiment to constrain the non-unitarity parameters. In order to do this, we fix the true value of $\delta_{CP}$ in  its currently preferred value $-\pi/2$ and assume that the hierarchy of neutrino is normal, then simulate the true event spectra by assuming unitary mixing and compare it with test event spectra by assuming non-unitary mixing. The values of $\chi^2$ are evaluated using the standard rules as described in GLoBES and the details are presented in Appendix A.
While doing the analysis, we do marginalization over $\delta_{CP}$ and $\theta_{23}$.  We show the allowed regions for non-unitarity parameters in $|\eta_{21}|-\delta_{CP}$ ($\phi_{21}-\delta_{CP}$)  plane in the left (middle) panel of Fig.\ref{allowedn}. From the figure, we can see that non unitarity parameters are sensitive to NO$\nu$A experiment. It can  also be seen from the figure that the 1$\sigma$, 2$\sigma$, and 3$\sigma$  contours are around the $-\pi/2$ as expected and there is a chance of degenerate solution at higher C.L. 
 
Furthermore, we would like to see the discovery reach of non-unitary parameter $|\eta_{21}|$ at NO$\nu$A experiment. We test the non-unitary mixing against the unitary mixing as mentioned before and  also do marginalization over true values of $\delta_{CP}$. The obtained sensitivity as a function of $\eta_{21}$ is shown in the right panel of  Fig. \ref{allowedn}. It can be inferred from the figure that the parameter space allowed by NO$\nu$A experiment at 1$\sigma$ C.L. is $|\eta_{21}|< 0.033$, which is a weaker constraint on this parameter compared to  the constraints obtained in other oscillation physics searches. Therefore, NO$\nu$A experiment is not expected to improve the current knowledge of non-unitarity parameter $\eta_{21}$.
 \subsection{Impact of non-unitarity parameters on the determination of unknowns}
In this subsection, we discuss how the unitarity violation in lepton mixing affect the sensitivity of neutrino mass hierarchy, octant of atmospheric mixing angle and leptonic CP violating phase $\delta_{CP}$. The degeneracies among the oscillation parameters play crucial role in the determination of these unknowns. Therefore, we also discuss how the degeneracies among the oscillation parameters get affected in presence of non-unitary mixing. 

\begin{figure}
\begin{center}
\minipage{0.25\textwidth}
    \includegraphics[height=5cm,width=7cm]{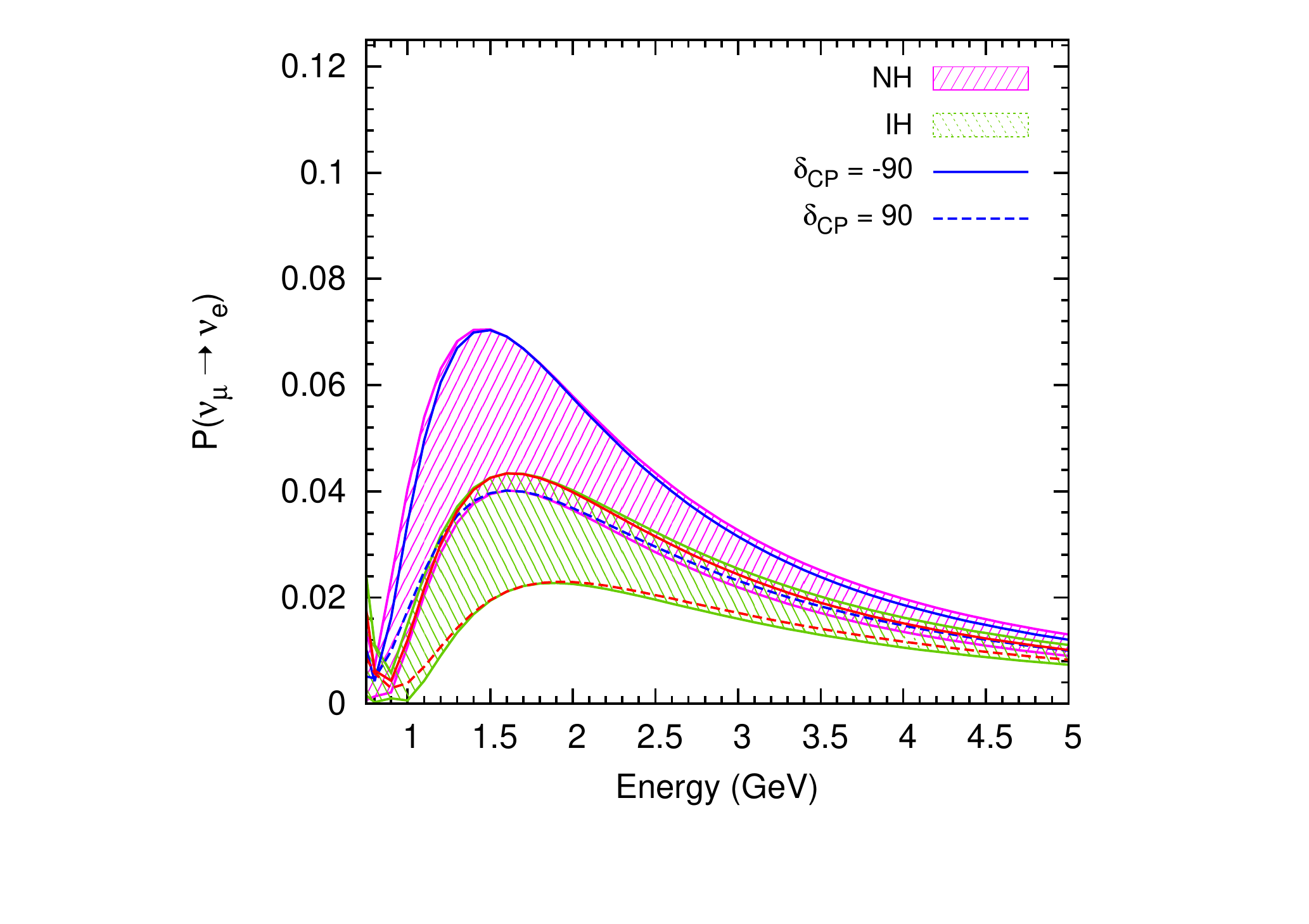}    \endminipage\hfill
\minipage{0.25\textwidth}
    \includegraphics[height=5cm,width=7cm]{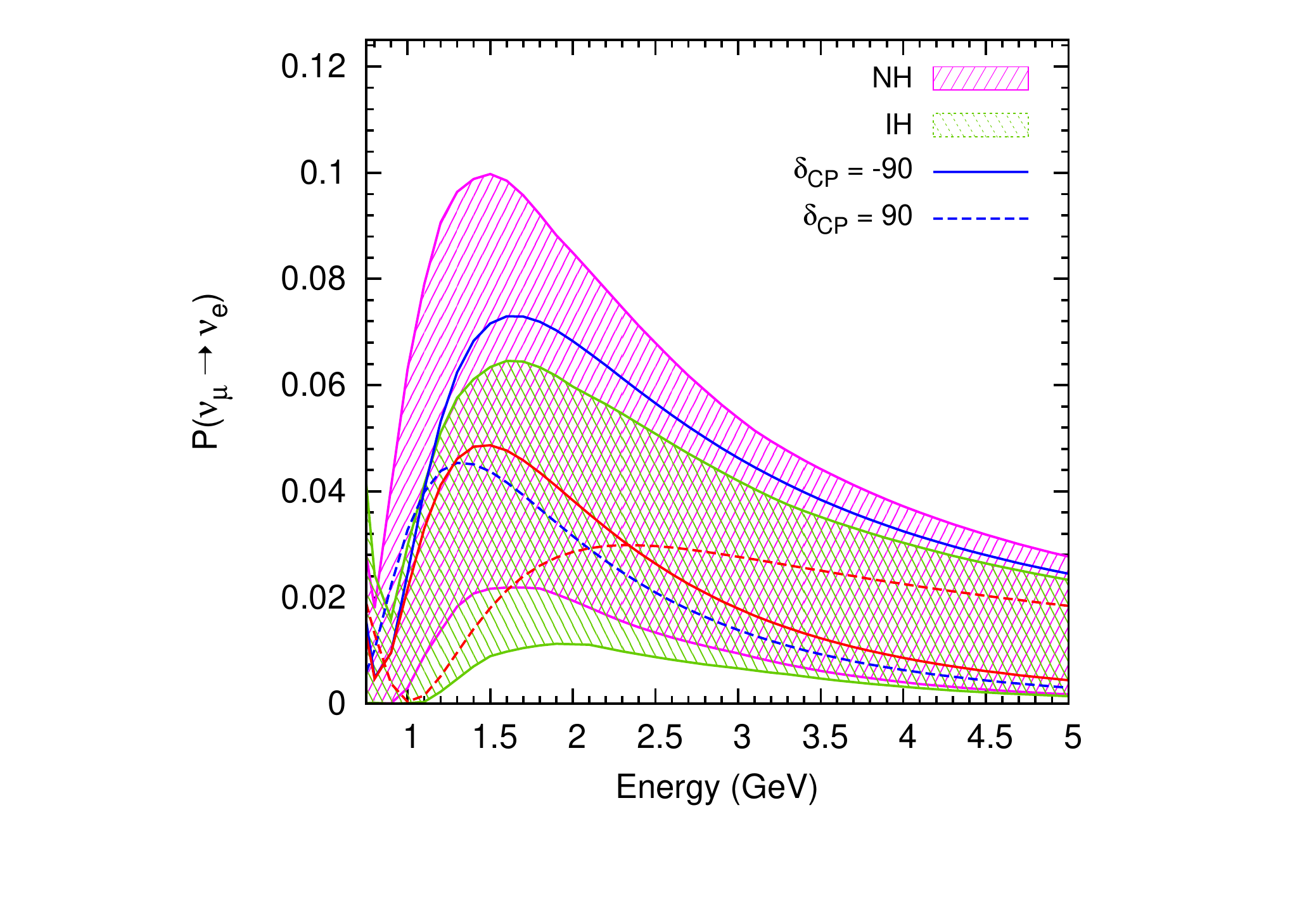}    \endminipage\hfill
\minipage{0.35\textwidth}
    \includegraphics[height=5cm,width=7cm]{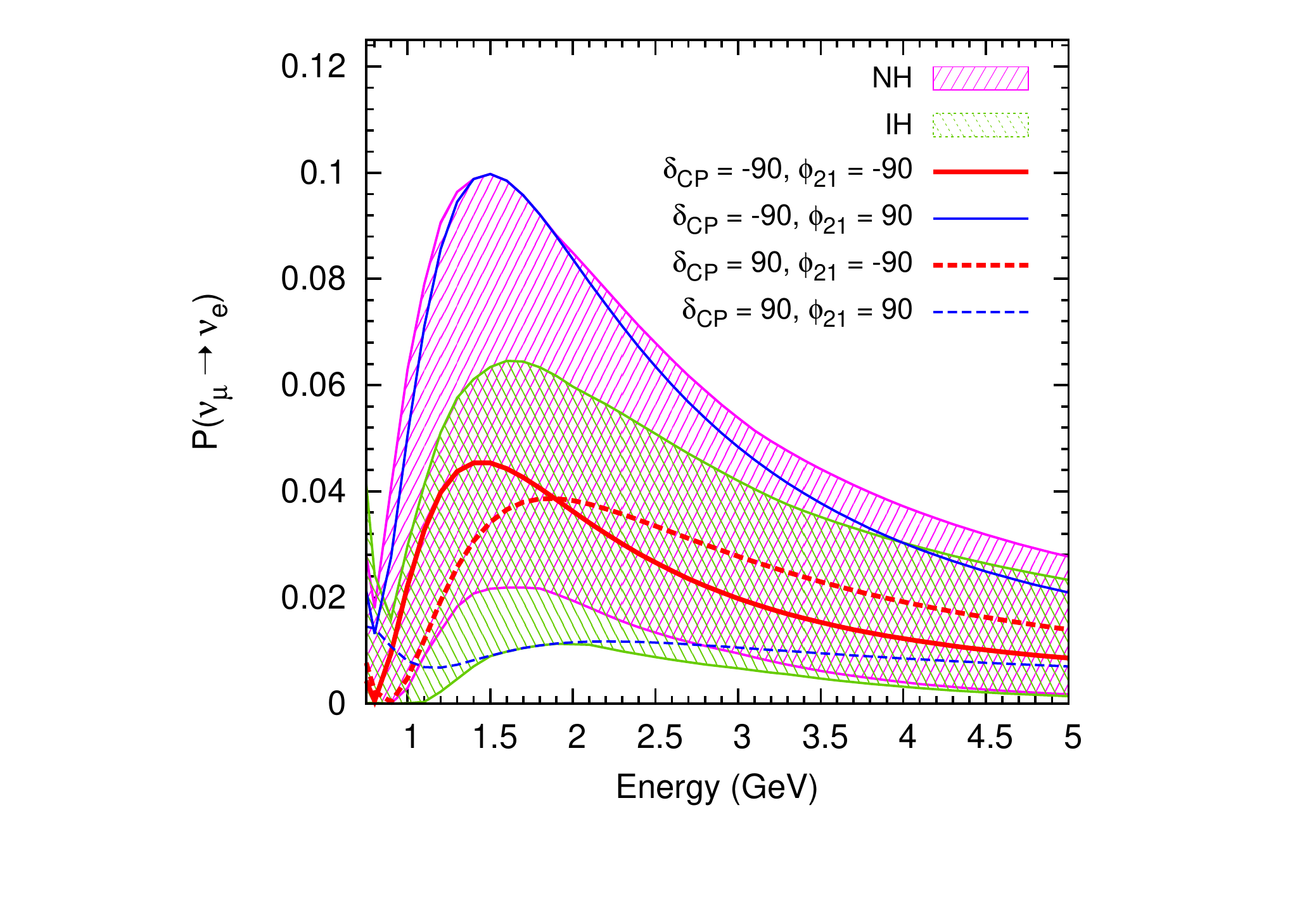}
    \endminipage
  \end{center}
\caption{{\label{mh-osc}} The oscillation probability  as a function of energy. The left (middle/right) panel corresponds to  oscillation without (with) unitarity violation in lepton mixing.  In the left and middle panels, the bands are obtained by varying $\delta_{CP}$ in its allowed range. Whereas in the right panel the bands are obtained by varying both $\delta_{CP}$ and $\phi_{21}$ in their allowed values. The solid (dashed) curve corresponds to oscillation probability for $\delta_{CP} = -90^\circ~ (90^\circ)$. The thick (thin)  curve in the right panel  corresponds to $\phi_{21} = -90^\circ~ (90^\circ)$.}
\end{figure}

In the left panel of Fig. \ref{mh-osc}, we show the oscillation probability for $\delta_{CP}$ in the range [$-\pi:\pi$] for  both normal (magenta band) and inverted (green band) hierarchies. The overlapped region is due to the degeneracy between the CP-violating phase $\delta_{CP}$ and neutrino mass hierarchy. If the true value of $\delta_{CP}$ lies in the overlapped region, then it is difficult to determine the mass hierarchy of neutrino. Whereas, the values of $\delta_{CP}$ far away from the overlapped regions can determine the mass hierarchy. From the left panel of the figure, we can see that the solid (dashed) curve in the NH (IH) band is for $\delta_{CP}= -90^\circ~ (90^\circ)$, which lies far away from the overlapped region. Therefore, $[-\pi:0]~([0:\pi])$ is the favourable region for the normal (inverted) mass hierarchy.  However, in the presence of non-unitary mixing there exists more overlapping between the NH and IH as one can see from the middle panel of the figure. In this case, the $\delta_{CP}= -90^\circ~( 90^\circ)$ curve is  also laying near to (within) the overlapped regions which results in the deterioration of MH sensitivity. If we invoke the phase $(\phi_{21})$ contribution of the non-unitary mixing, then we end up with a case as shown in the right panel of figure. From this figure, it is clear that the $\delta_{CP}= -90^\circ ~(90^\circ)$ with $\phi_{21}= 90^\circ~(-90^\circ)$ is favourable for the determination of NH (IH) hierarchy as it  lies far away from the overlapped region. 
While doing this analysis we assume that
$\theta_{23}=45^{\circ}$. 

\begin{figure}
 \begin{center}
  \includegraphics[height=6cm,width=8cm]{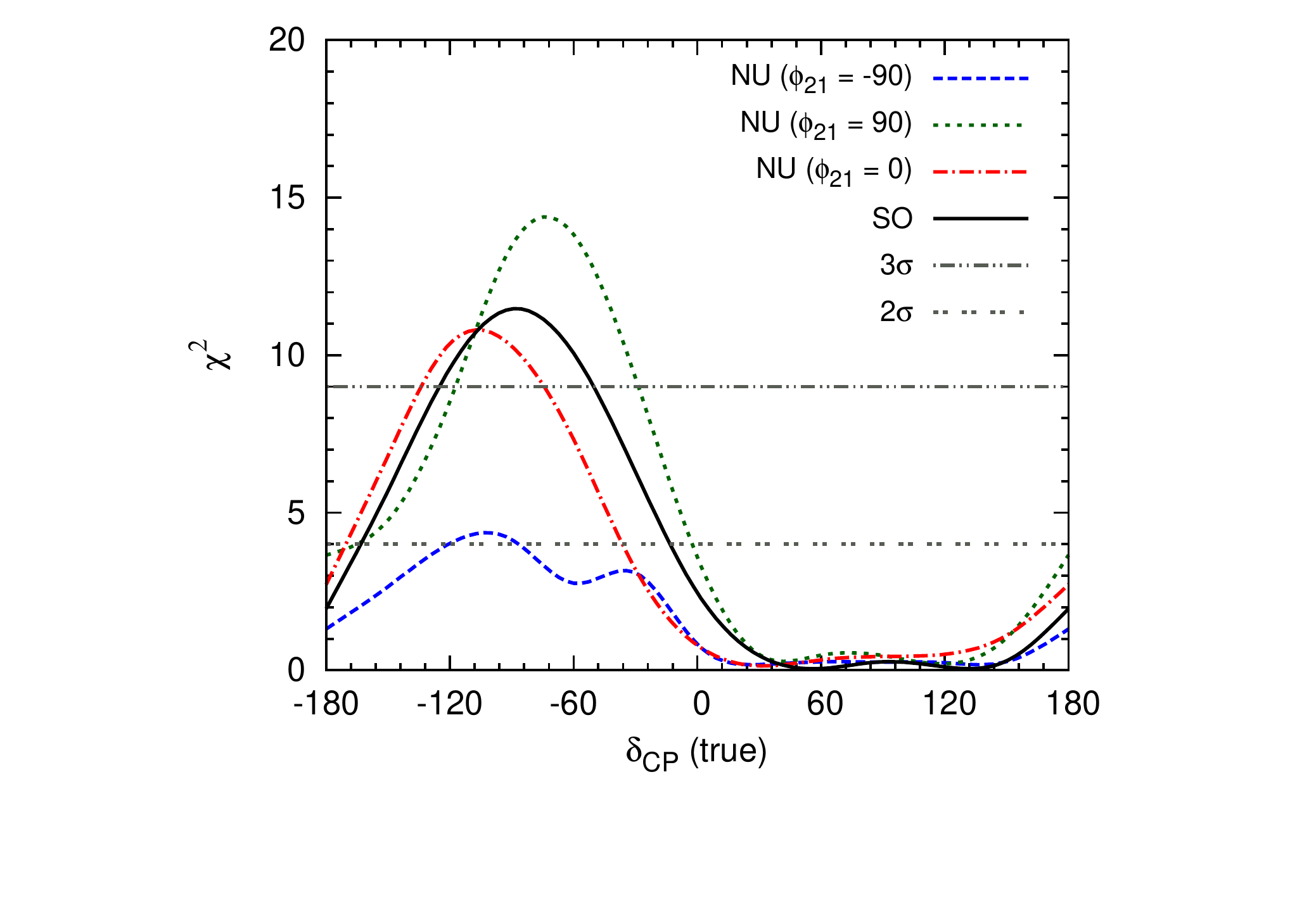}
 \includegraphics[height=6cm,width=8cm]{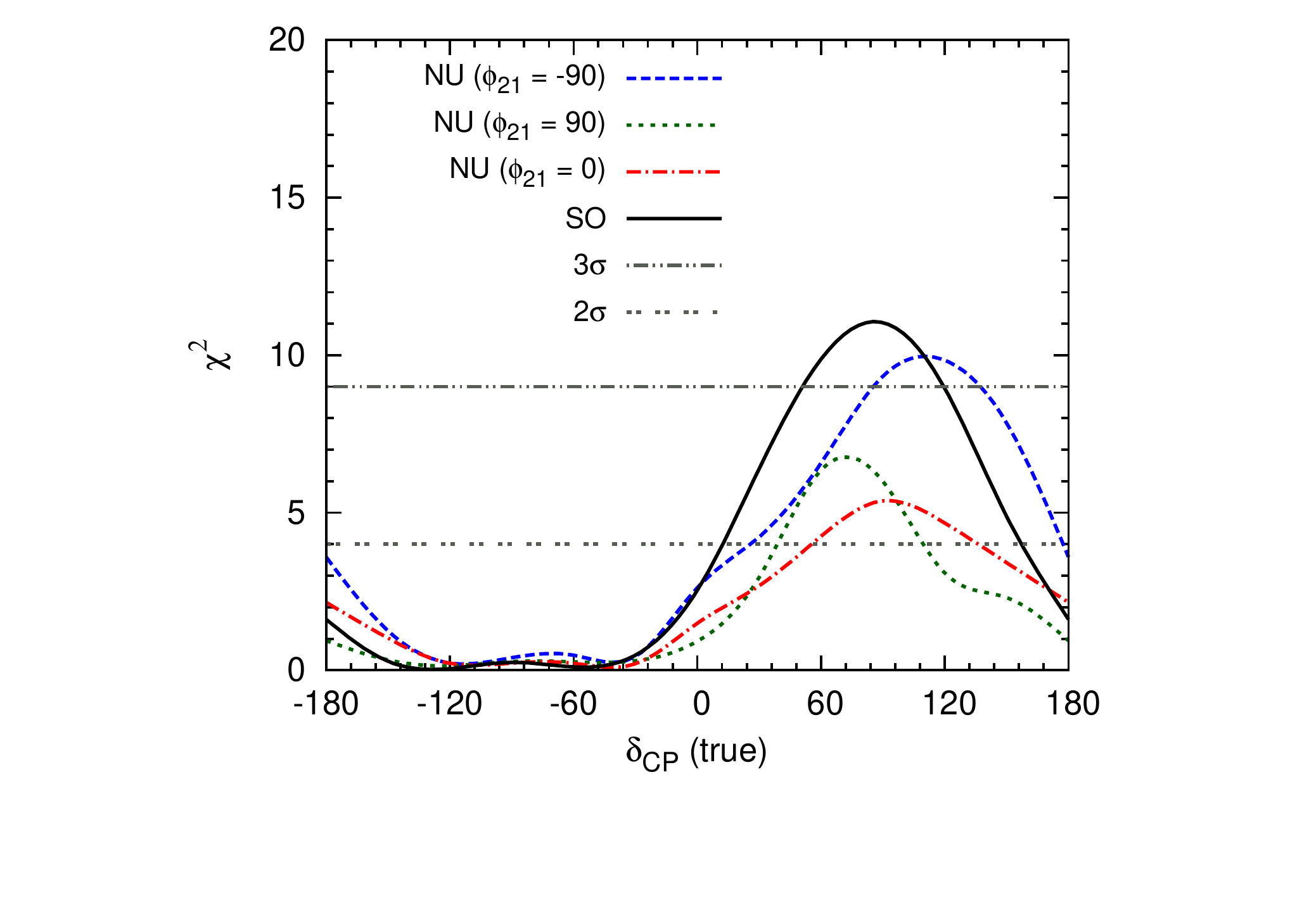}
 \caption{{\label{mh-sens}}The MH-sensitivity for NO$\nu$A. The true hierarchy is assumed to be  normal (inverted) in the left (right) panel.}
 \end{center}
 \end{figure}

Further, we show the MH sensitivity in Fig. \ref{mh-sens}. To obtain the MH sensitivity, we assume that the true hierarchy is normal (inverted) and do comparison between  the true event spectra and the test event spectra with inverted (normal)  hierarchy. While doing the analysis, we do marginalisation over $\delta_{CP}$, $\theta_{23}$ and $\phi_{21}$ in their allowed 3$\sigma$ ranges.  The obtained sensitivity as a function of true value of $\delta_{CP}$ is shown in the left (right) panel of the figure, where the true hierarchy is assumed to be normal (inverted). 
 From the left (right) panel of the figure, we can see that,  in the  standard oscillation framework, if the true mass hierarchy of neutrino is normal (inverted) and the true value of $\delta_{CP}$ is around $-90^\circ~ (90^\circ)$, then  
 it is possible to determine mass hierarchy at a C.L. above 3$\sigma$ by using NO$\nu$A experiment. 
   For non-unitary case, we show MH sensitivity for three different values of new phase $\phi_{21}= 0,~ 90^\circ,$ and $-90^\circ$. Though the sensitivity is reduced significantly in the presence of non-unitary parameter $\phi_{21} = 0,90^\circ$ ($\phi_{21}=0,-90^\circ$), there is a possibility that mass hierarchy can be determined  with more than 
3$\sigma$ C.L. if  the $\delta_{CP}$ lies around $-90^\circ~ (90^\circ)$ and the $\phi_{21}$ is around  $90^\circ~ (-90^\circ)$ for normal (inverted) hierarchy.
  
Another important open question related to neutrino oscillation physics is the determination of octant of atmospheric mixing angle. The oscillation data show that atmospheric neutrino mixing is not maximal, which implies that $\theta_{23}$ can be either in Lower Octant or in Higher Octant. Moreover, recent experimental results on $\theta_{23}$ always show tension between these two octants. Therefore, it is quite important to study the sensitivity of octant in presence of non-unitary lepton mixing. 

\begin{figure}
\begin{center}
\minipage{0.25\textwidth}
    \includegraphics[height=5cm,width=7cm]{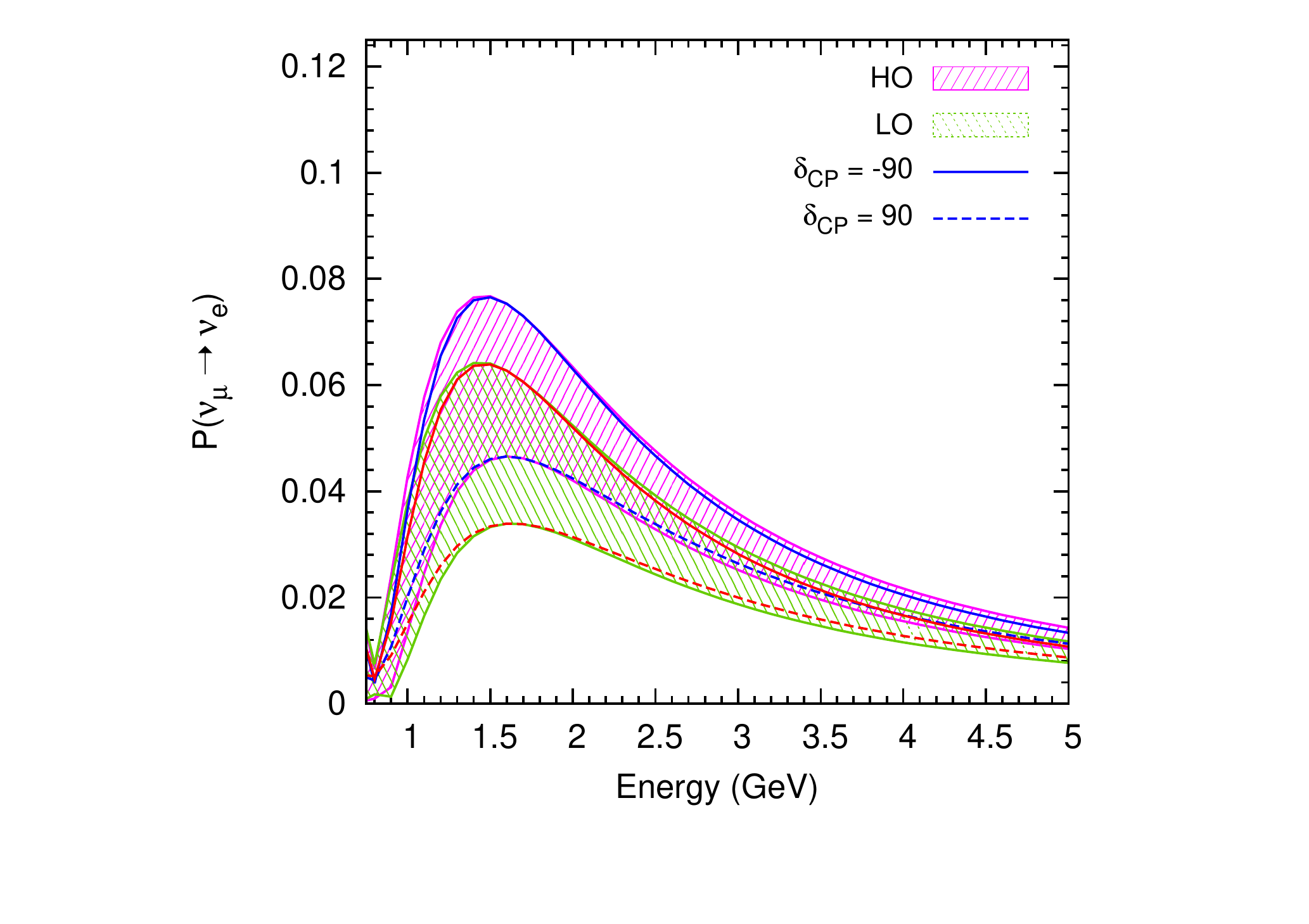}    \endminipage\hfill
\minipage{0.25\textwidth}
    \includegraphics[height=5cm,width=7cm]{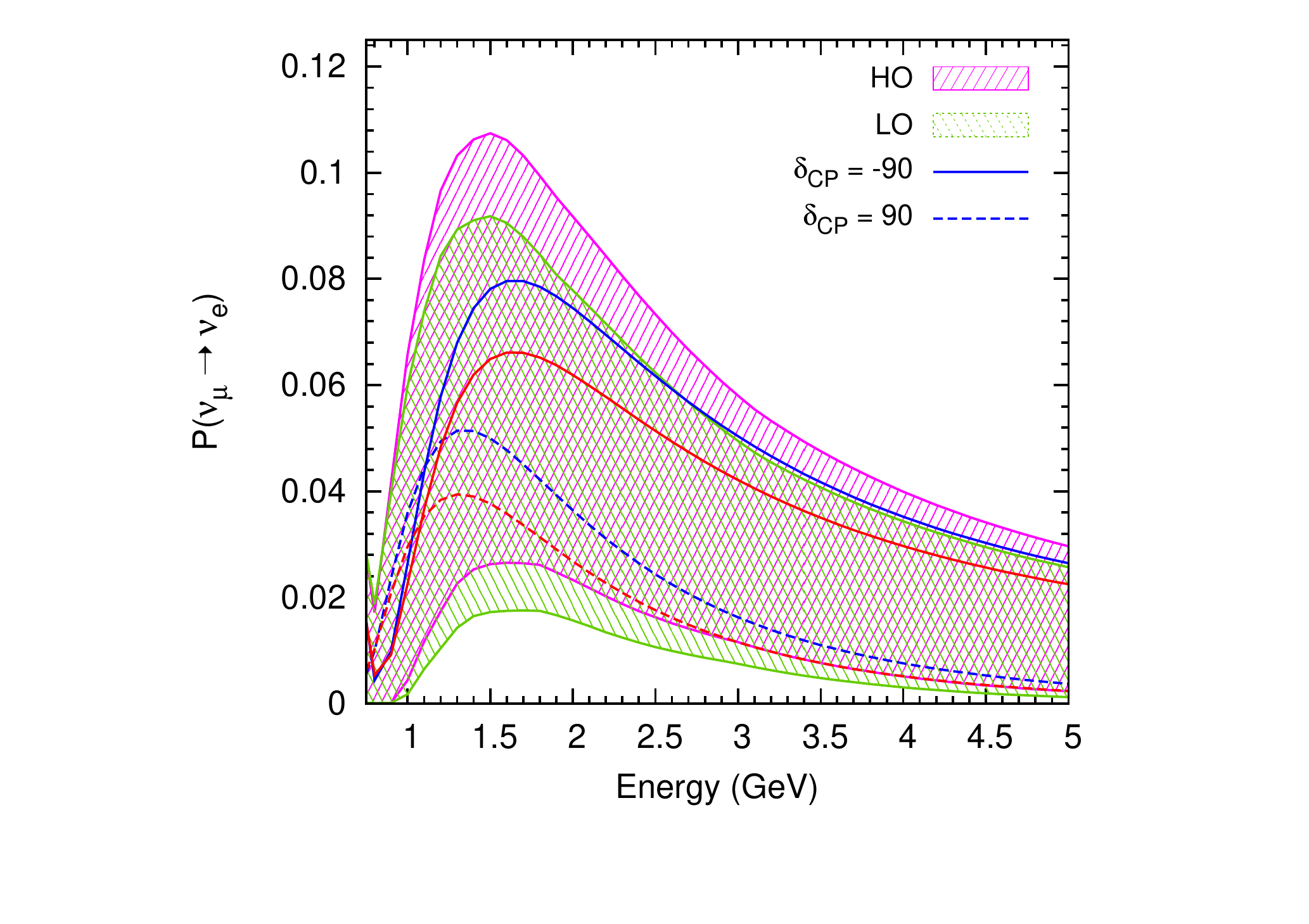}    \endminipage\hfill
\minipage{0.35\textwidth}
    \includegraphics[height=5cm,width=7cm]{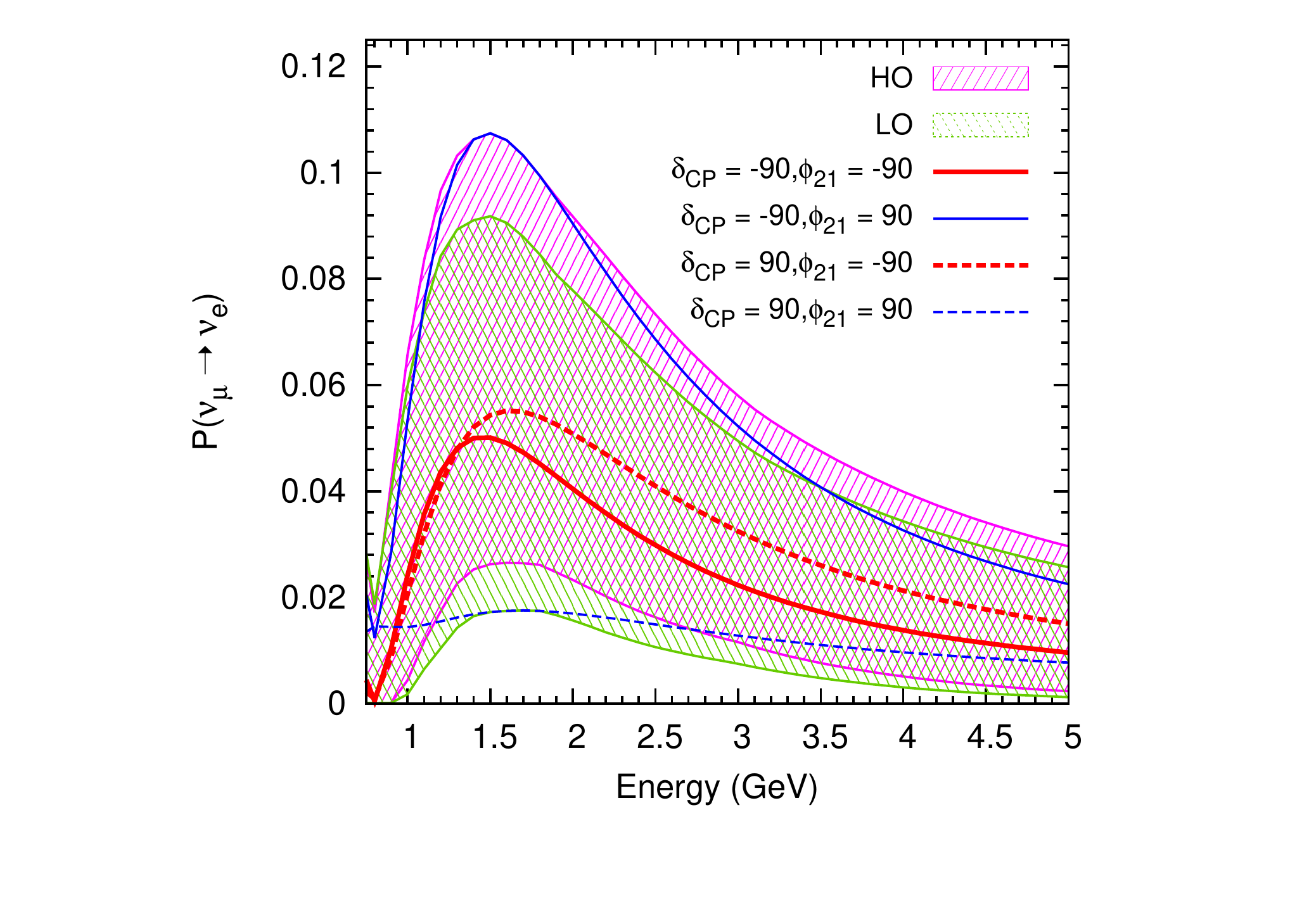}
    \endminipage
  \end{center}
\caption{{\label{octant-osc}} The oscillation probability  as a function of energy. The left (middle/right) panel corresponds to  oscillation without (with) unitarity violation in lepton mixing.  In the left and middle panels, the bands are obtained by varying $\delta_{CP}$ in its allowed range. Whereas in the right panel the bands are obtained by varying both $\delta_{CP}$ and $\phi_{21}$ within their allowed ranges. The solid (dashed) curve corresponds to oscillation probability for $\delta_{CP} = -90^\circ~ (90^\circ)$. The thick (thin)  curve in the right panel  corresponds to $\phi_{21} = -90^\circ~ (90^\circ)$.}
\end{figure}

In the left panel of Fig.\ref{octant-osc}, we show the oscillation probability for $\delta_{CP}$ in the range ($-\pi:\pi$) for  both HO (magenta band) and LO (green band) by assuming that the mass
hierarchy of neutrino is normal. The overlapped region is due to the degeneracy between the CP-violating phase $\delta_{CP}$ and atmospheric mixing angle $\theta_{23}$. If the true value of $\delta_{CP}$ lies in the overlapped region, then it is difficult to determine the octant of  $\theta_{23}$. Whereas, the values of $\delta_{CP}$ far away from the overlapped region can determine the octant of $\theta_{23}$. From the left panel of the figure, we can see that the solid (dashed) curve in the HO (LO) band is for $\delta_{CP}= -90^\circ~ (90^\circ)$, which lies far away from the overlapped region. Therefore, in the standard oscillation framework, $[-\pi:0]~([0:\pi])$ is the favourable region for the Higher (Lower) Octant.  However, in the presence of non-unitary mixing there exists more overlapping between the HO and LO as one can see from the middle panel of the figure, which results in the deterioration of octant sensitivity. If we invoke the phase $(\phi_{21})$ contribution of the non-unitary mixing, then we end up with a case as shown in the right panel of figure. From this figure, it is clear that  $\delta_{CP}= -90^\circ ~(90^\circ)$ with $\phi_{21}= 90^\circ~(-90^\circ)$ is favourable for the determination of HO (LO) as it is laying far away from the overlapped region.

 \begin{figure}
 \begin{center}
  \includegraphics[height=6cm,width=8cm]{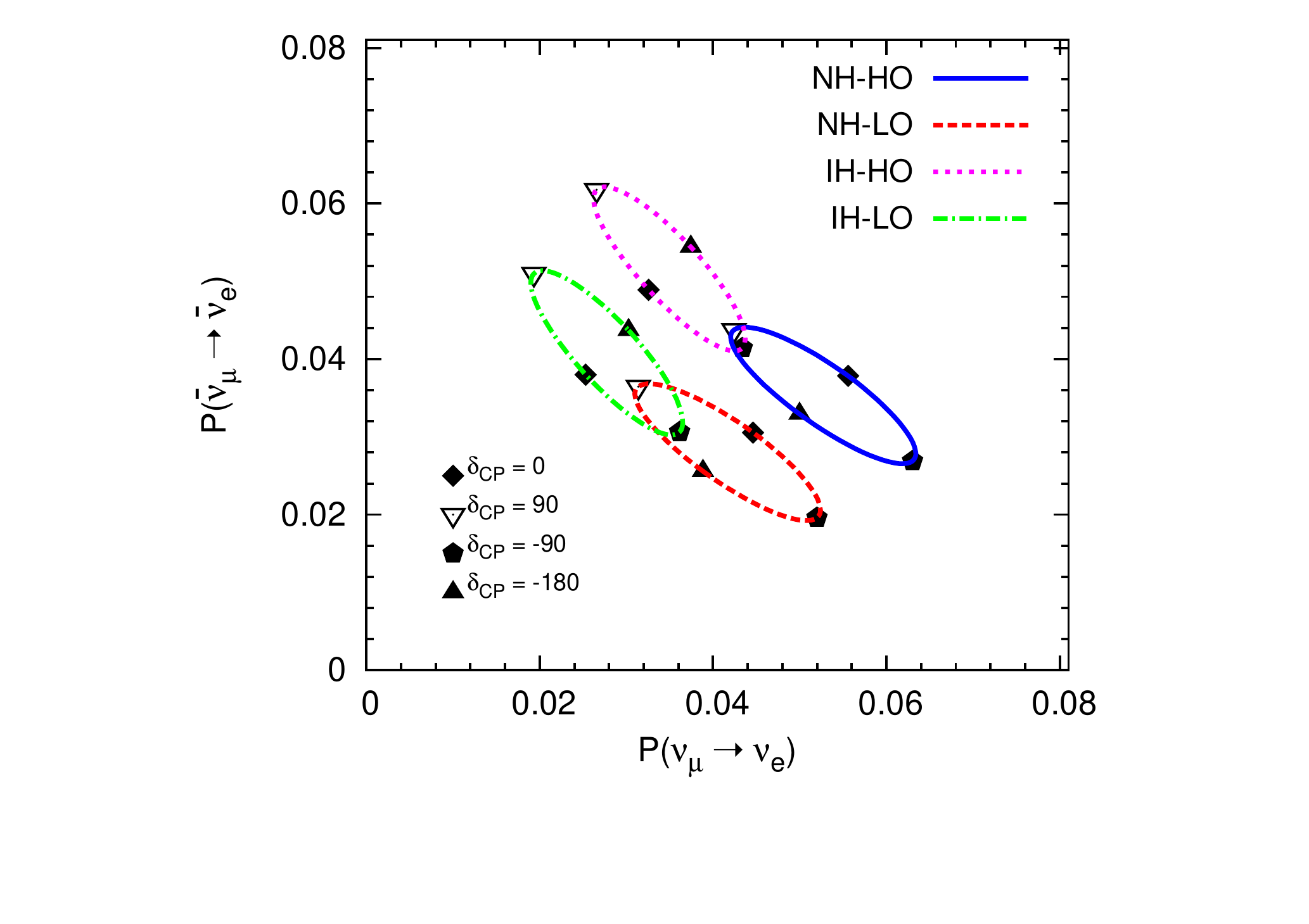}
 \includegraphics[height=6cm,width=8cm]{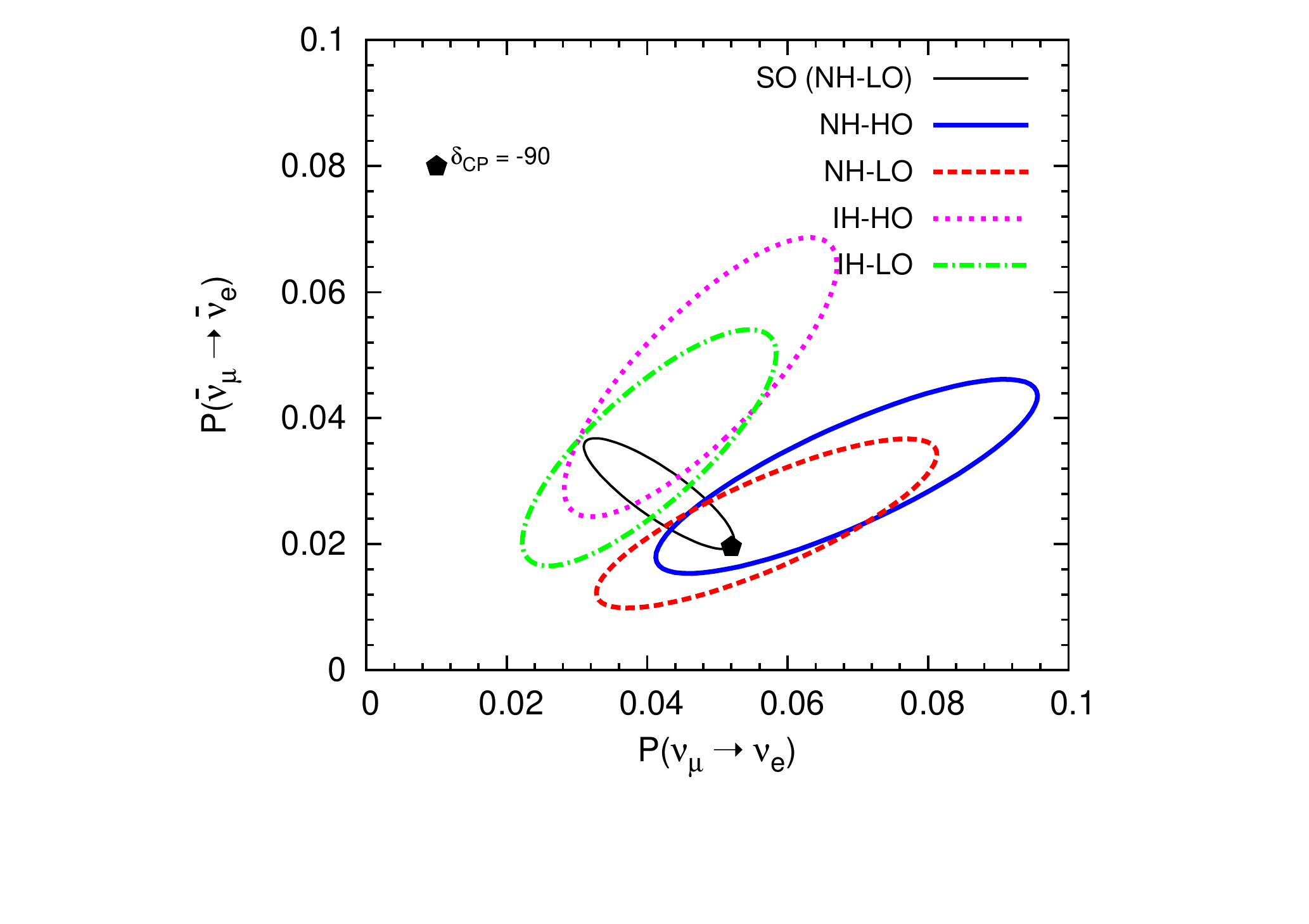}
 \caption{{\label{biporb}}The bi-probability plots for NO$\nu$A (L = 810 km, E = 2 GeV)  without (with) unitarity violation is in the left (right) panel.}
 \end{center}
 \end{figure}
One of the most convenient ways to demonstrate the existing degeneracies among the oscillation parameters (mass hierarchy, octant and $\delta_{CP}$ degeneracies)  is by using bi-probability curves, which show the oscillation probabilities for all possible values of CP-violating phase with fixed mass hierarchy and octant combinations in a neutrino-antineutrino oscillation probability plane as shown in  Fig. \ref{biporb}. In the standard paradigm of neutrino oscillation, there are mainly four degeneracies among the oscillation parameters NH-LO, NH-HO, IH-LO, and IH-HO, which give rise to four ellipses in the     $P(\nu_{\mu} \to \nu_e)$-$ P(\bar{\nu}_{\mu} \to \bar{\nu}_{e})$ plane as shown in the left panel of Fig. \ref{biporb}. From the figure, it can be seen that the ellipses for lower octant and higher octant are well separated, which indicates that NO$\nu$A can determine octant of atmospheric mixing angle. Whereas, the ellipses for normal hierarchy and inverted hierarchy
are overlapped with each other, especially in the case of lower octant.
It should be noted from the figure that the CP-violating phases $\delta_{CP}= 90^\circ,~ -90^\circ$ are laying far away from the overlapped regions. Therefore, if CP-phase is around these values, then it is possible to resolve octant and mass hierarchy degeneracies to a great extent. However, in presence of non-unitary mixing new CP-violating phase also comes into picture. Therefore, we obtain the ellipses by fixing $\delta_{CP}= -90^\circ$  and varying the phase of non-unitarity parameter ($\phi_{21}$). The thin solid (black) ellipse in the right panel of the figure  corresponds to LO-NH case in standard neutrino oscillation, which helps for a  direct comparison of unitary and non-unitary cases. It can be seen from the figure that the non-unitary lepton mixing leads to new degeneracies among the oscillation parameters which worsen the degeneracy resolution capability of NO$\nu$A experiment. 

\begin{figure}
 \begin{center}
  \includegraphics[height=6cm,width=8cm]{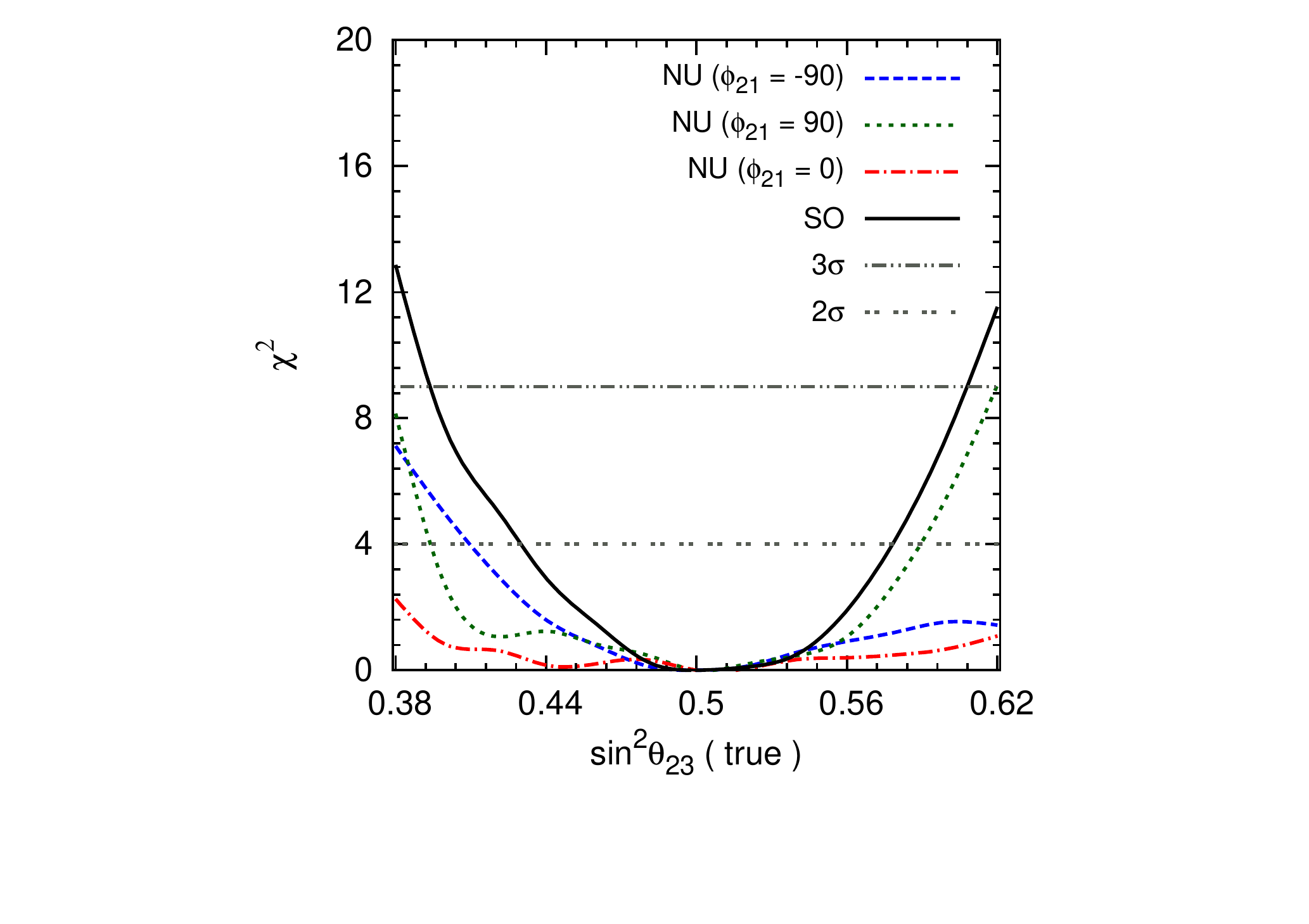}
 \includegraphics[height=6cm,width=8cm]{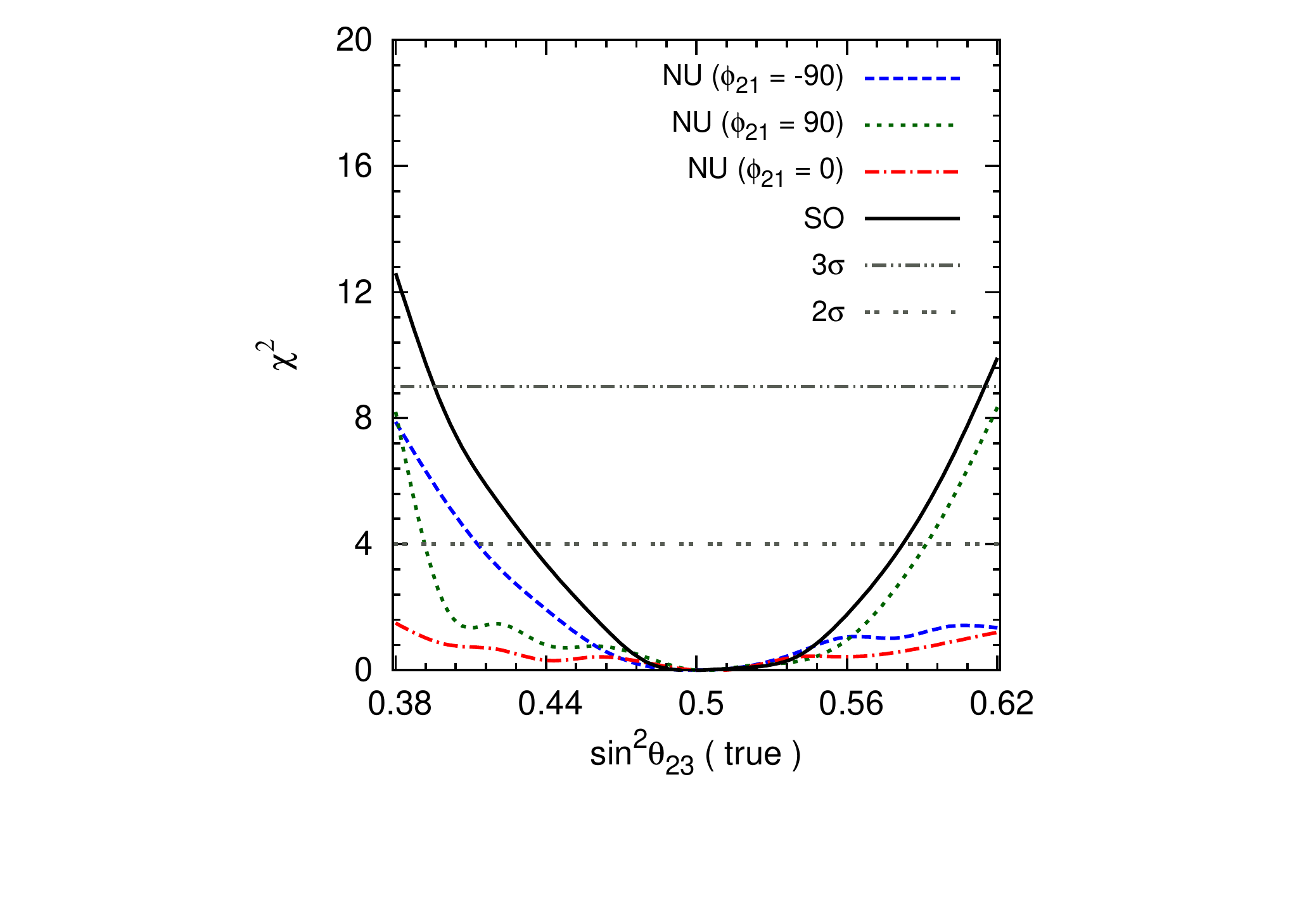}
 \caption{{\label{octant-ses}}The octant sensitivity for NO$\nu$A. The true hierarchy is assumed to be  normal (inverted) in the left (right) panel.}
 \end{center}
 \end{figure}

Next, we show the octant sensitivity of NO$\nu$A in Fig. \ref{octant-ses}. In order to calculate the sensitivity, we assume that the true octant of $\theta_{23}$ is HO (LO) and do a comparison between  the true event spectra and the test event spectra with LO (HO). While doing the analysis, we consider hierarchy to be normal and true value of $\delta_{CP}=-90^\circ$, and we do marginalisation over $\delta_{CP}$, $\phi_{21}$ in their allowed 3$\sigma$ ranges and $\theta_{23}$ in its allowed LO (HO) range.  The obtained sensitivity as a function of true value of $\sin^2\theta_{23}$ is shown in the figure. The hierarchy is assumed to be normal (inverted) in left (right)  panel of the figure. For non-untary case, we show  the octant sensitivity with three different values of new phase $\phi_{21}=0,90^\circ$ and $-90^\circ$. From the figure, it can be seen that if nature prefers a LO (HO) for $\theta_{23}$ with $\sin^2\theta_{23}$ = 0.41 (0.59), then the octant of $\theta_{23}$ can be determined at 2$\sigma$ C.L in the standard oscillation picture. However, the sensitivity is reduced in the case of non-unitary mixing with $\phi_{21}=0$. Though the sensitivity is significantly reduced   for $\phi_{21}= 90^\circ$ ($\phi_{21}=-90^\circ$) in the case of LO (HO) octant, there is a possibility that octant sensitivity can be determined in presence of non-unitary mixing if $\phi_{21}$ is around $-90^\circ$ ($90^\circ$) for LO (HO) as shown in the figure.

 \begin{figure}
 \begin{center}
  \includegraphics[height=6cm,width=8cm]{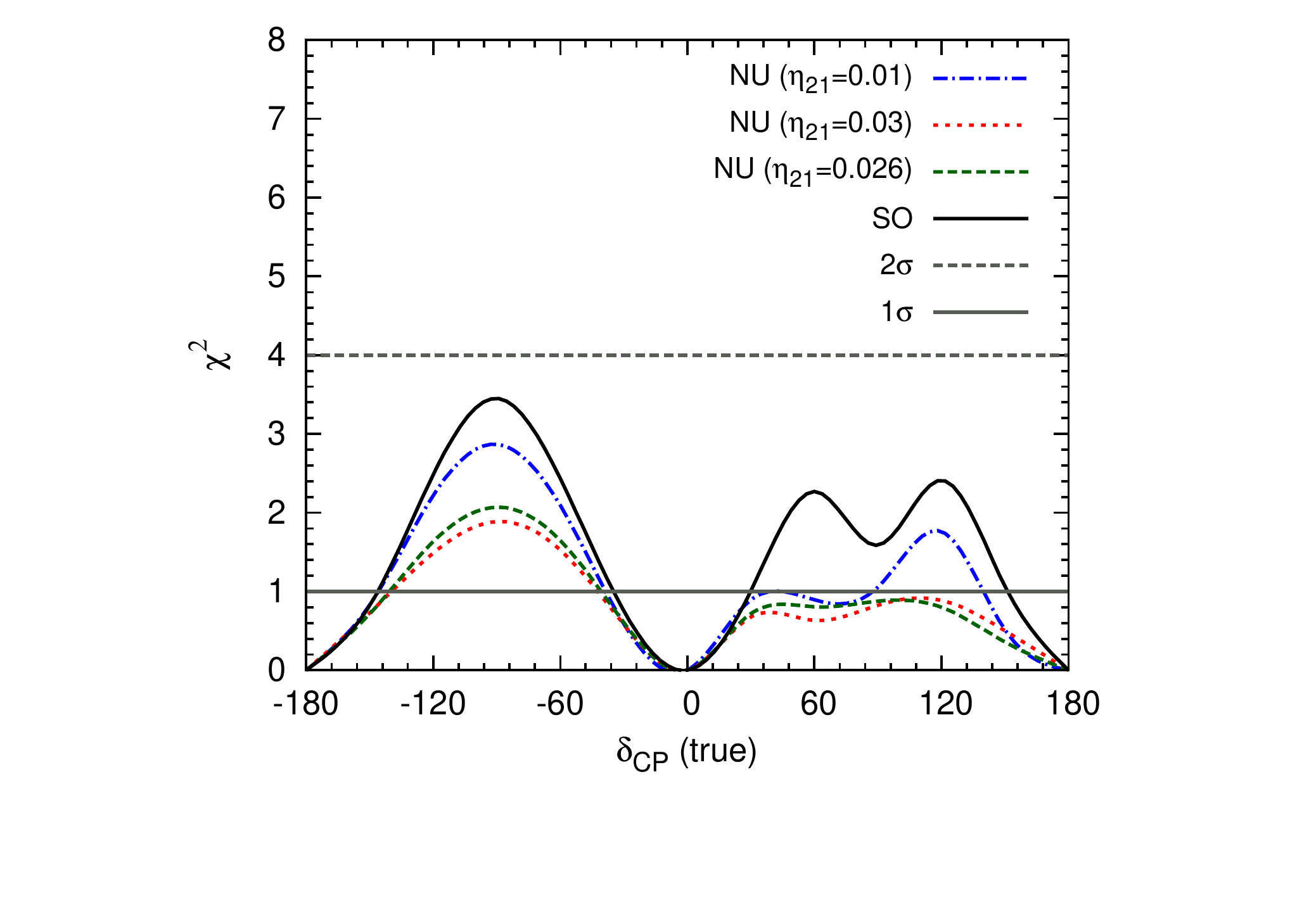}
 \includegraphics[height=6cm,width=8cm]{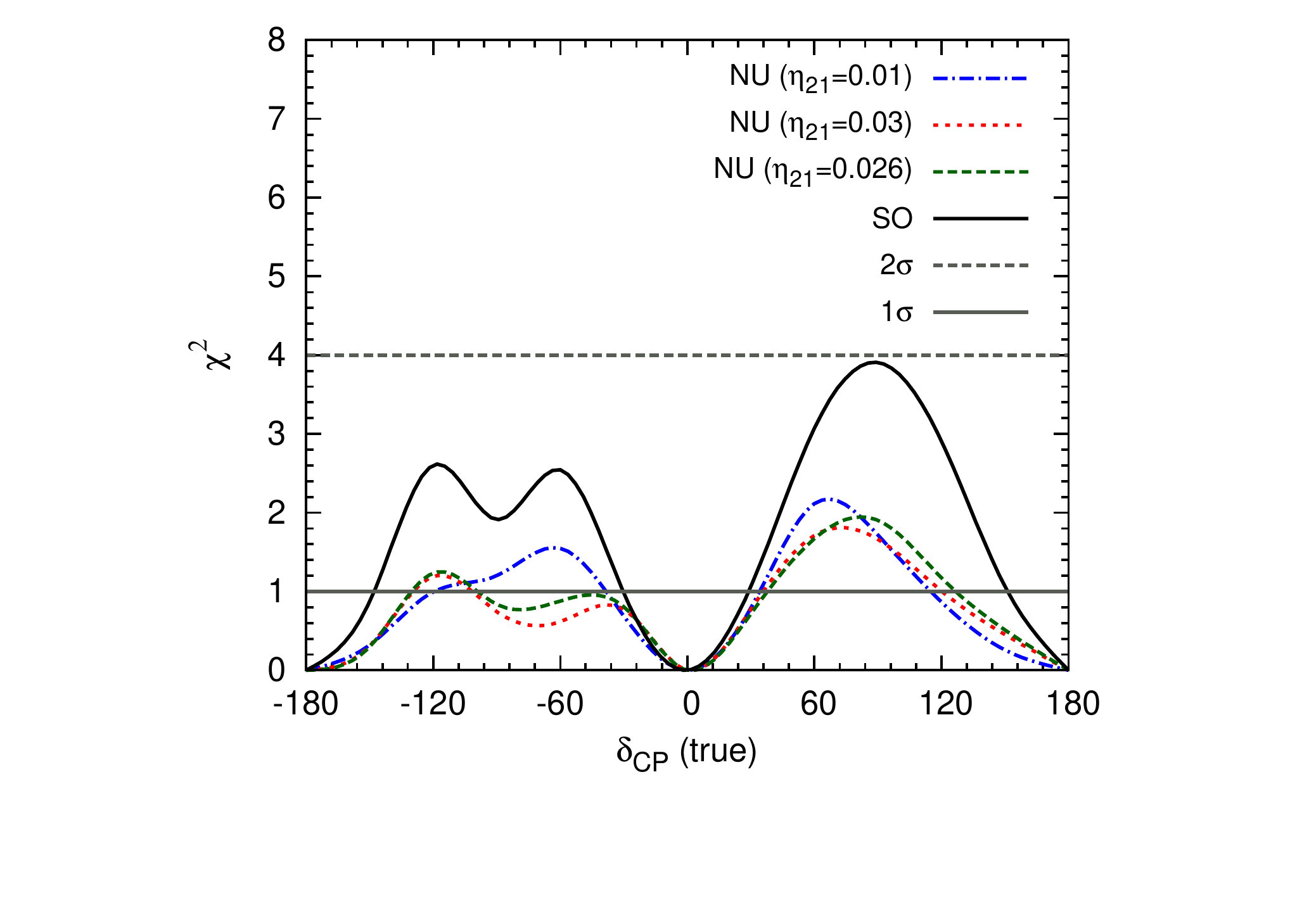}
 \caption{{\label{cpv-sens}}The CPV-sensitivity for NO$\nu$A. The true hierarchy is assumed to be  normal (inverted) in the left (right) panel.}
 \end{center}
 \end{figure}

Finally, we present the CPV sensitivity in Fig. \ref{cpv-sens}. To obtain the CPV sensitivity, we simulate the true event spectra  for each value of $\delta_{CP}$ and compare it against CP conserving test event spectra. This sensitivity is obtained for unknown mass hierarchy and marginalising over allowed values of $\theta_{23}$ and the non-unitarity parameter $\phi_{21}$. The obtained sensitivity as
a function of true $\delta_{CP}$ is shown in the figure.
 From the figure, it can be seen that in presence of non-unitary mixing the CPV sensitivity is reduced significantly.
\subsection{Degeneracy resolution of oscillation parameters in presence of non-unitarity mixing}
 \begin{figure}
 \begin{center}
  \includegraphics[height=5cm,width=8cm]{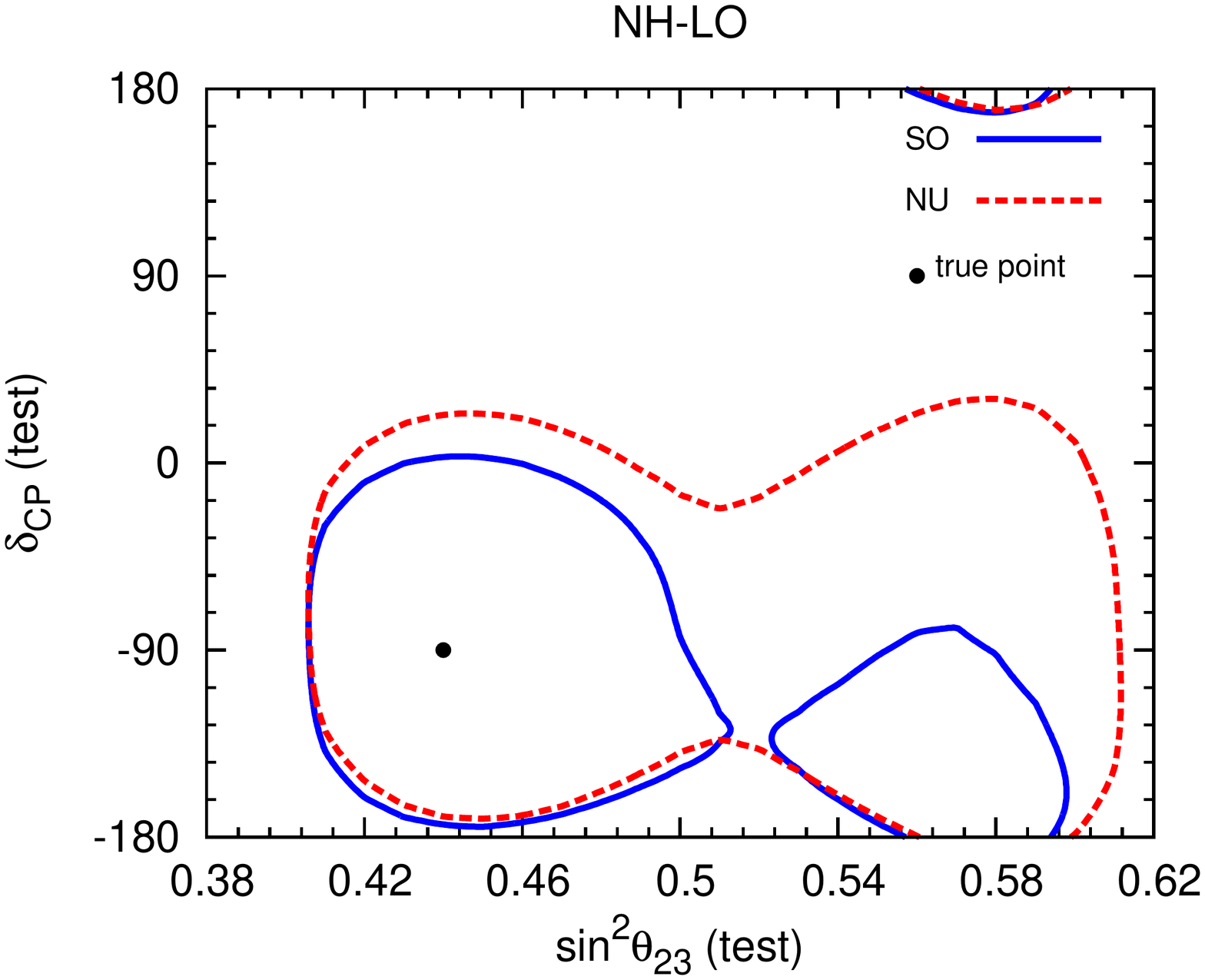}
  \includegraphics[height=5cm,width=8cm]{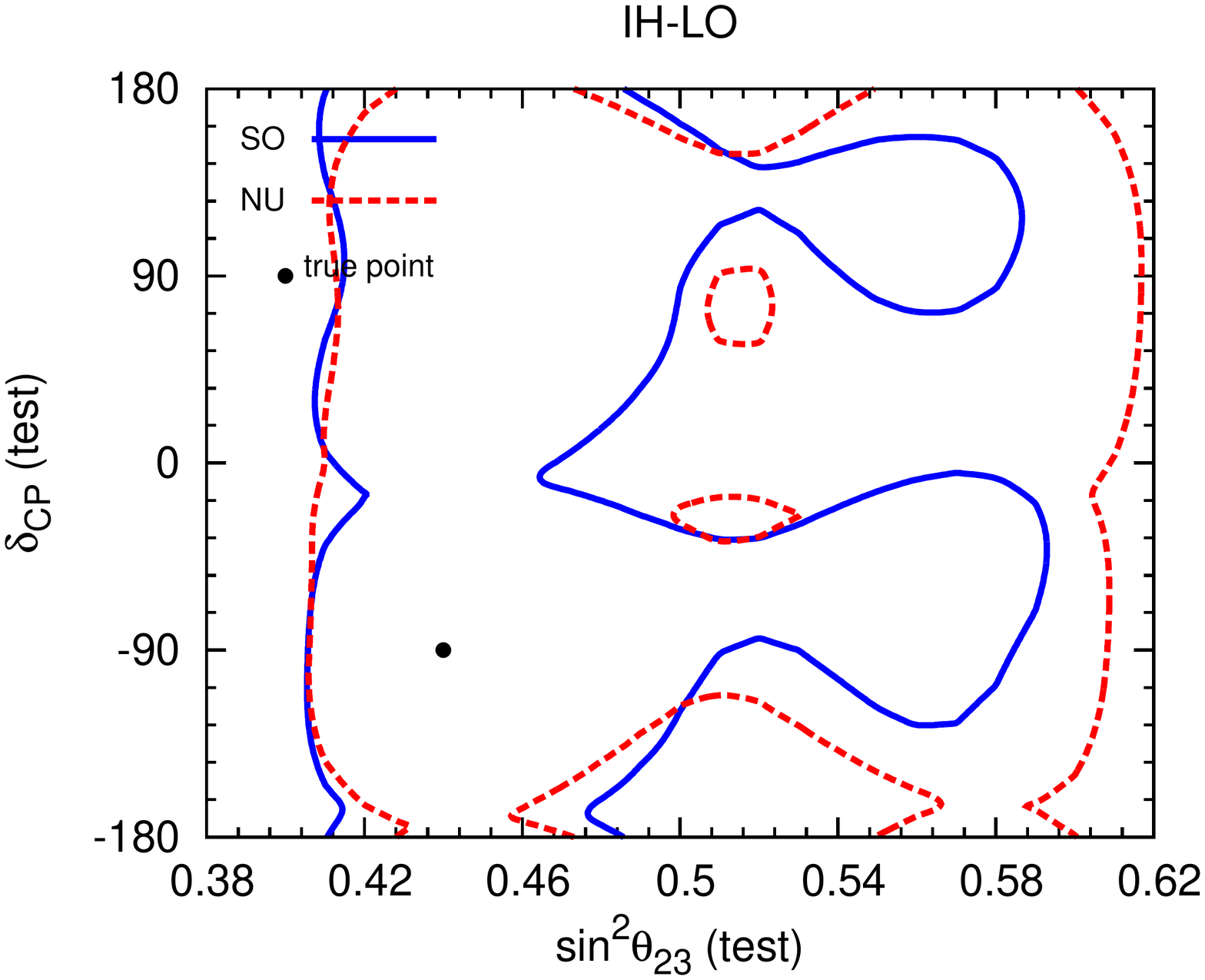}\\
  \includegraphics[height=5cm,width=8cm]{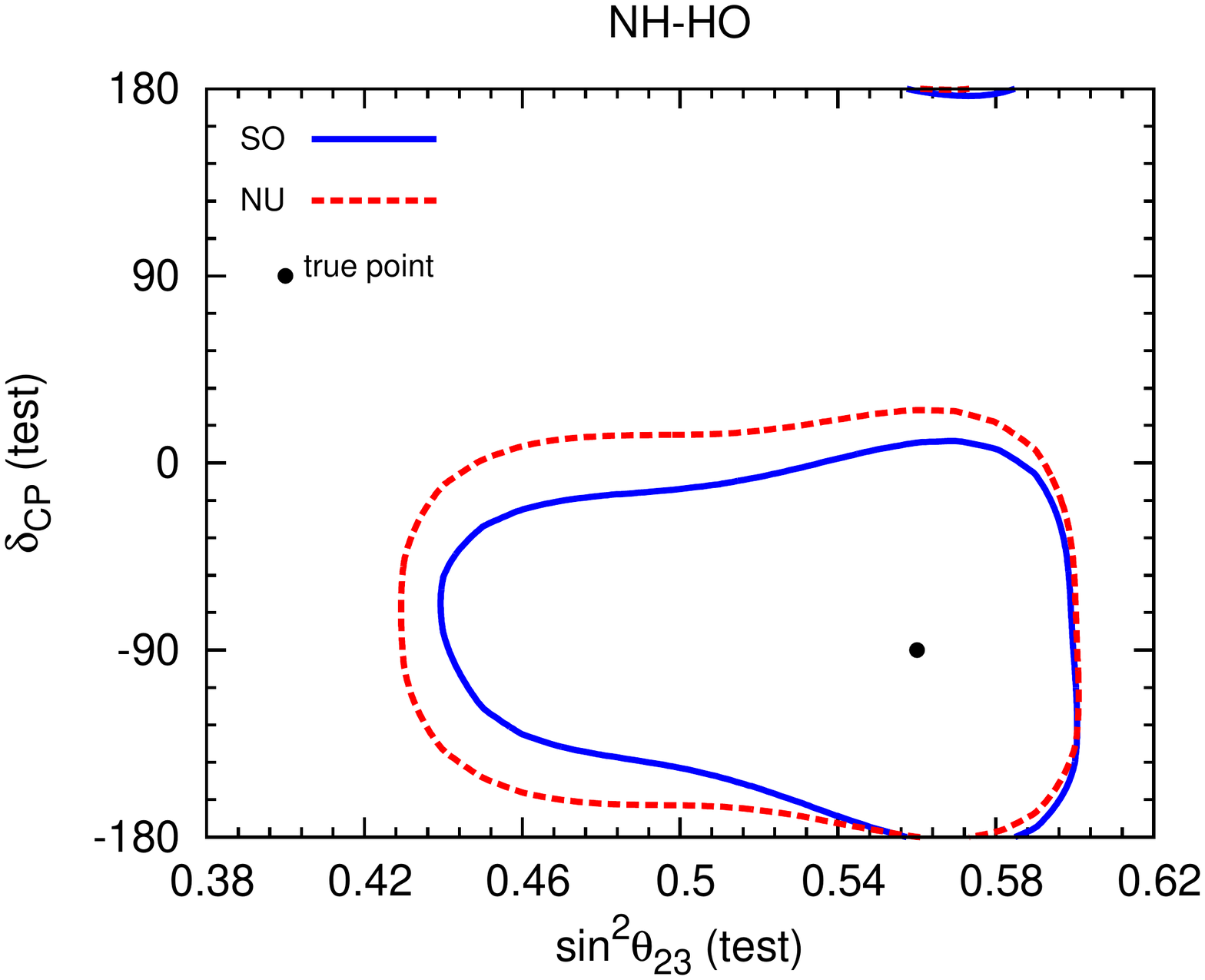}
  \includegraphics[height=5cm,width=8cm]{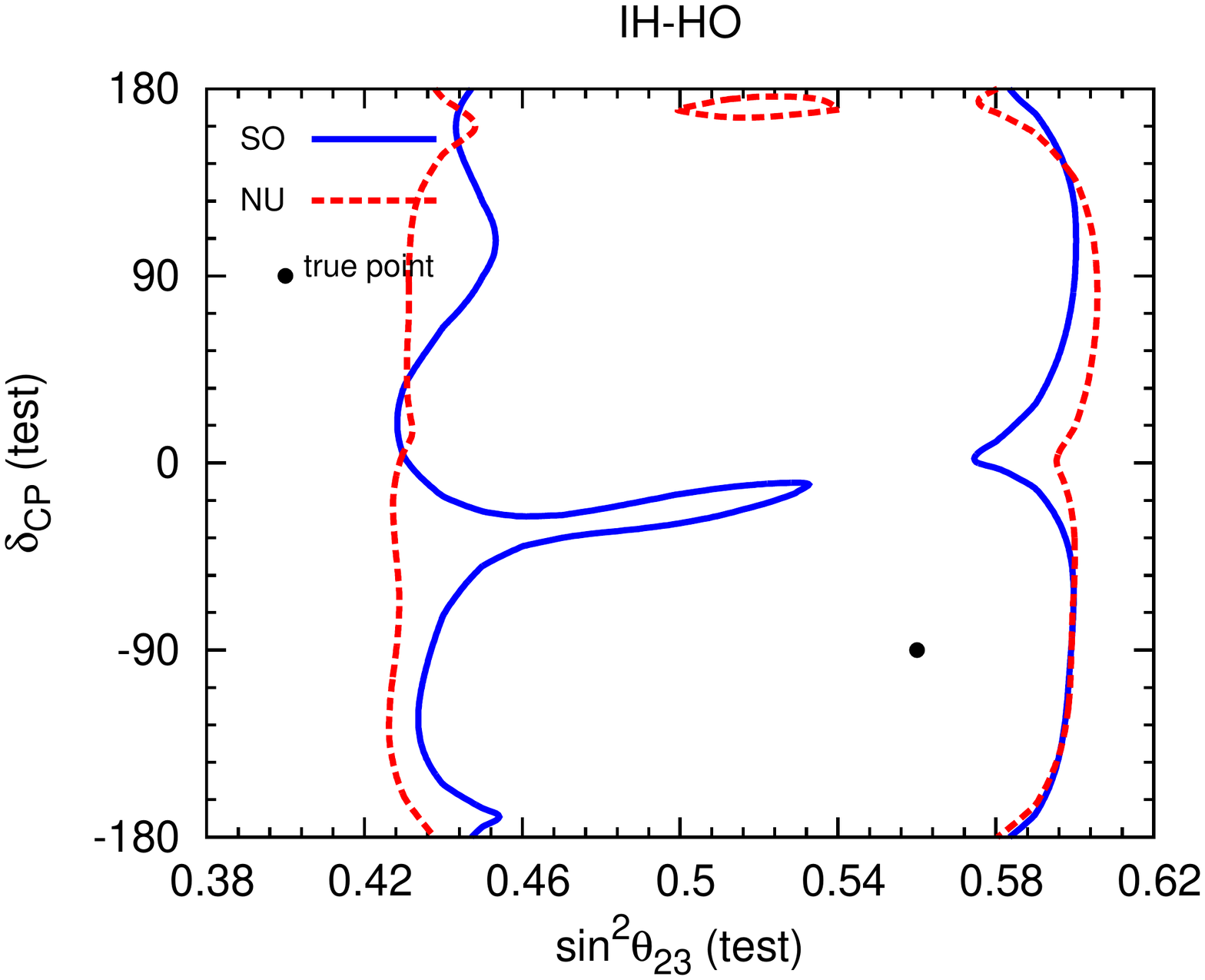}
 \caption{{\label{non-nova-allowed}}90\% C. L. allowed parameter space in $\sin^2\theta_{23}-\delta_{CP}$ plane for NO$\nu$A. The  hierarchy is assumed to be  normal (inverted) in the left (right) panel and octant of atmosphereic mixing angle is assumed to be LO (HO) in the top (bottom) panel.}
 \end{center}
 \end{figure}

\begin{figure}
 \begin{center}
  \includegraphics[height=5cm,width=8cm]{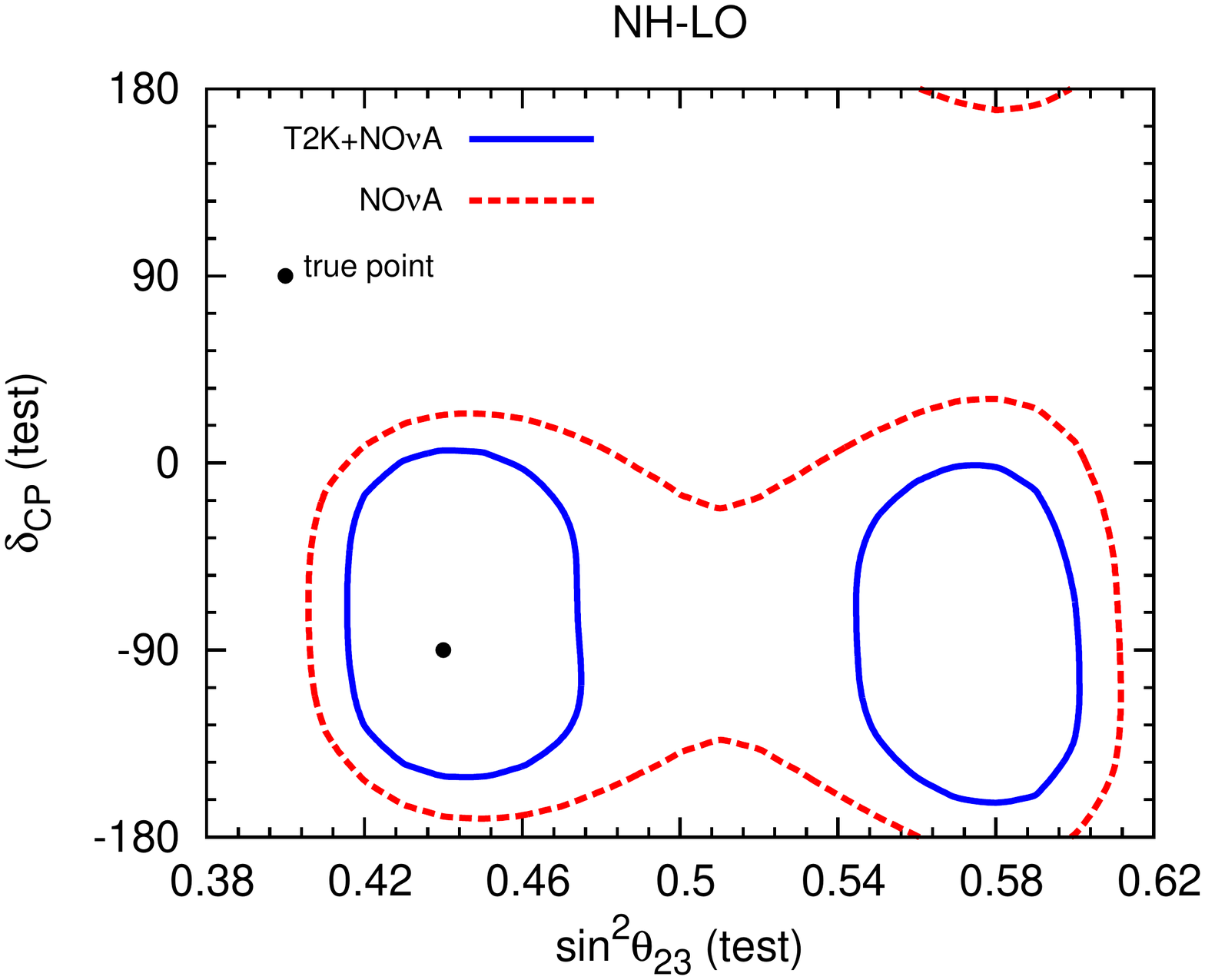}
  \includegraphics[height=5cm,width=8cm]{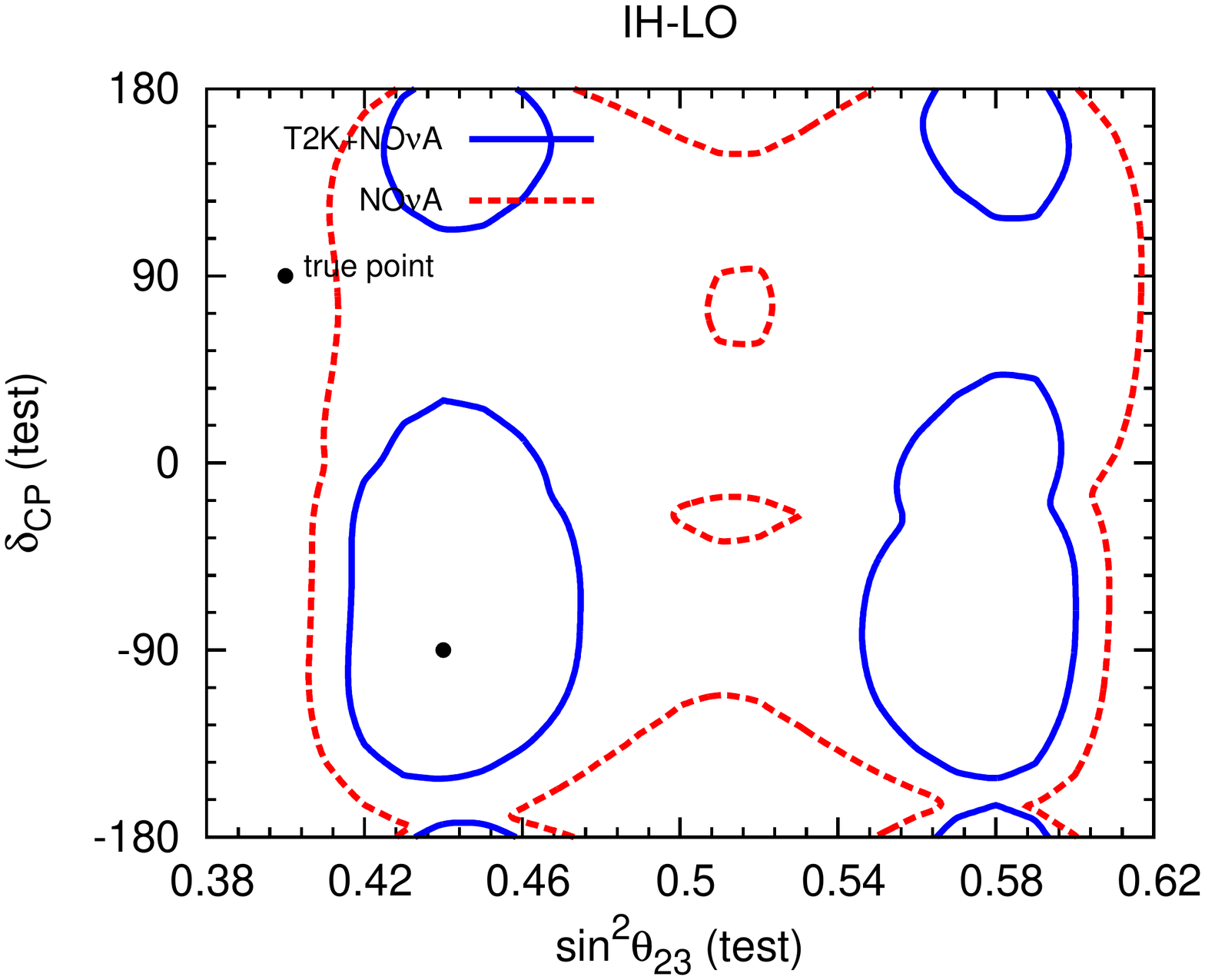}\\
  \includegraphics[height=5cm,width=8cm]{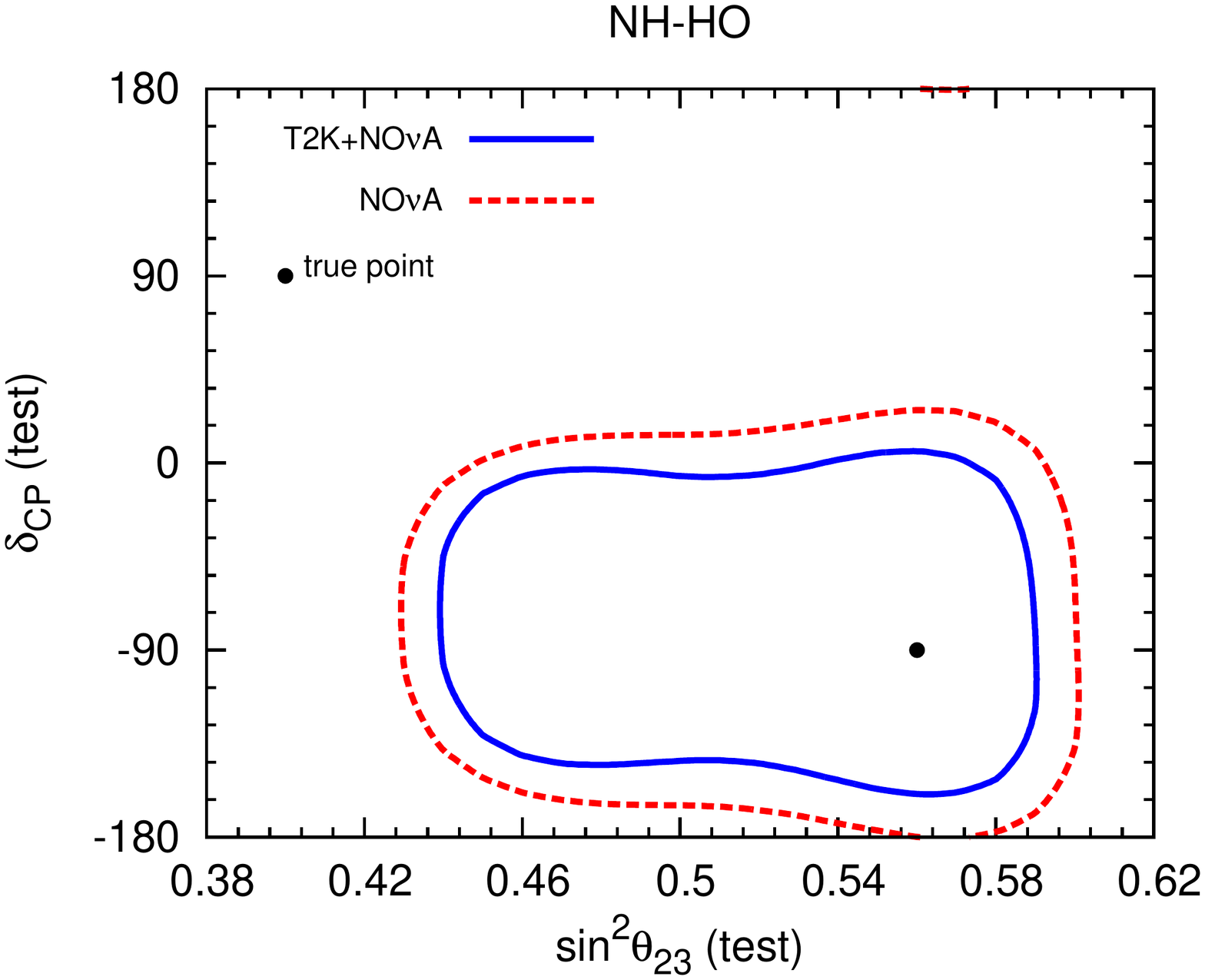}
  \includegraphics[height=5cm,width=8cm]{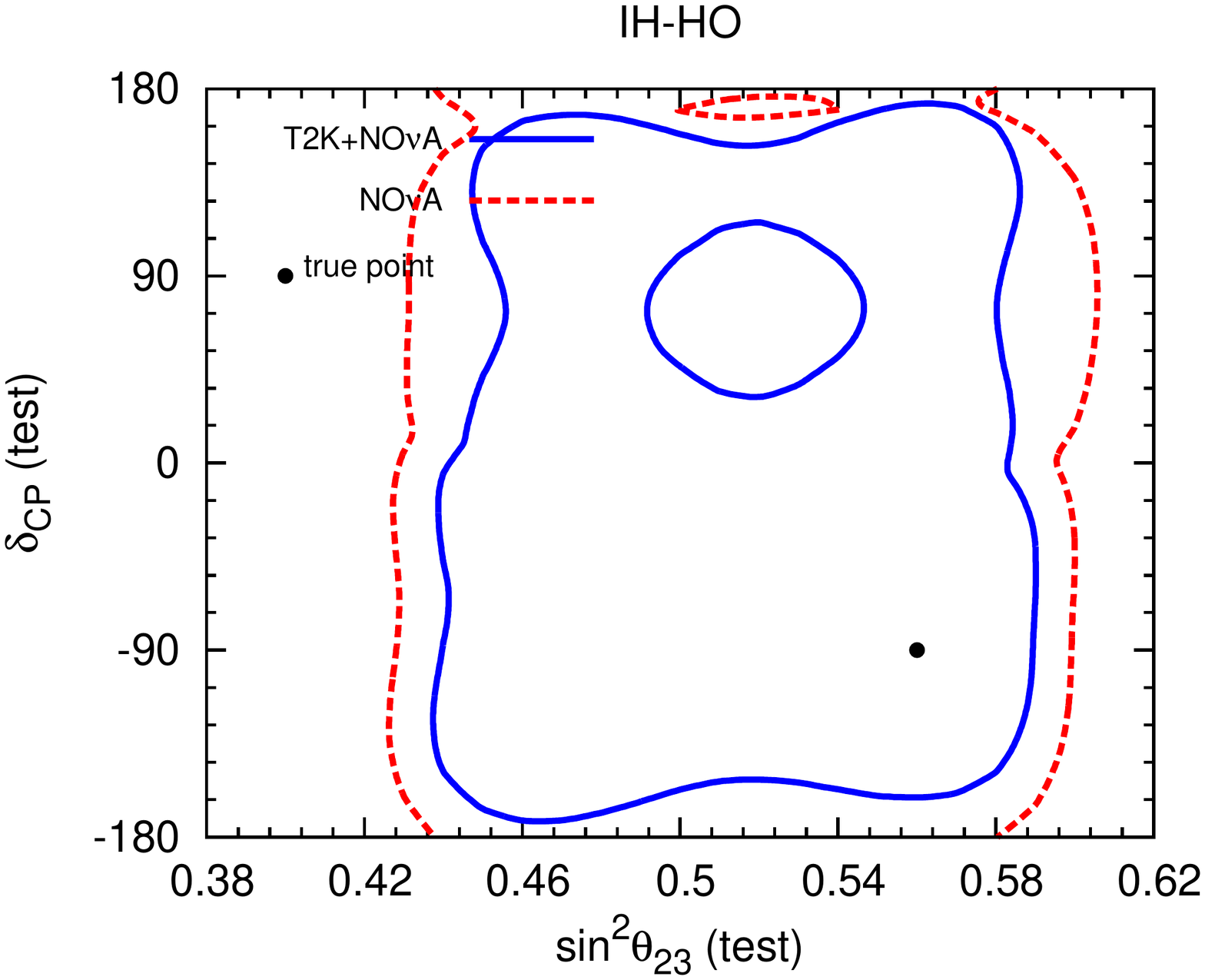}
 \caption{{\label{non-novat2k-allowed}}90\% C. L. allowed parameter space in $\sin^2\theta_{23}-\delta_{CP}$ plane for NO$\nu$A and T2K . The  hierarchy is assumed to be  normal (inverted) in the left (right) panel and octant of atmospheric mixing angle is assumed to be LO (HO) in the top (bottom) panel.}
 \end{center}
 \end{figure}

\begin{figure}
 \begin{center}
  \includegraphics[height=5cm,width=5cm]{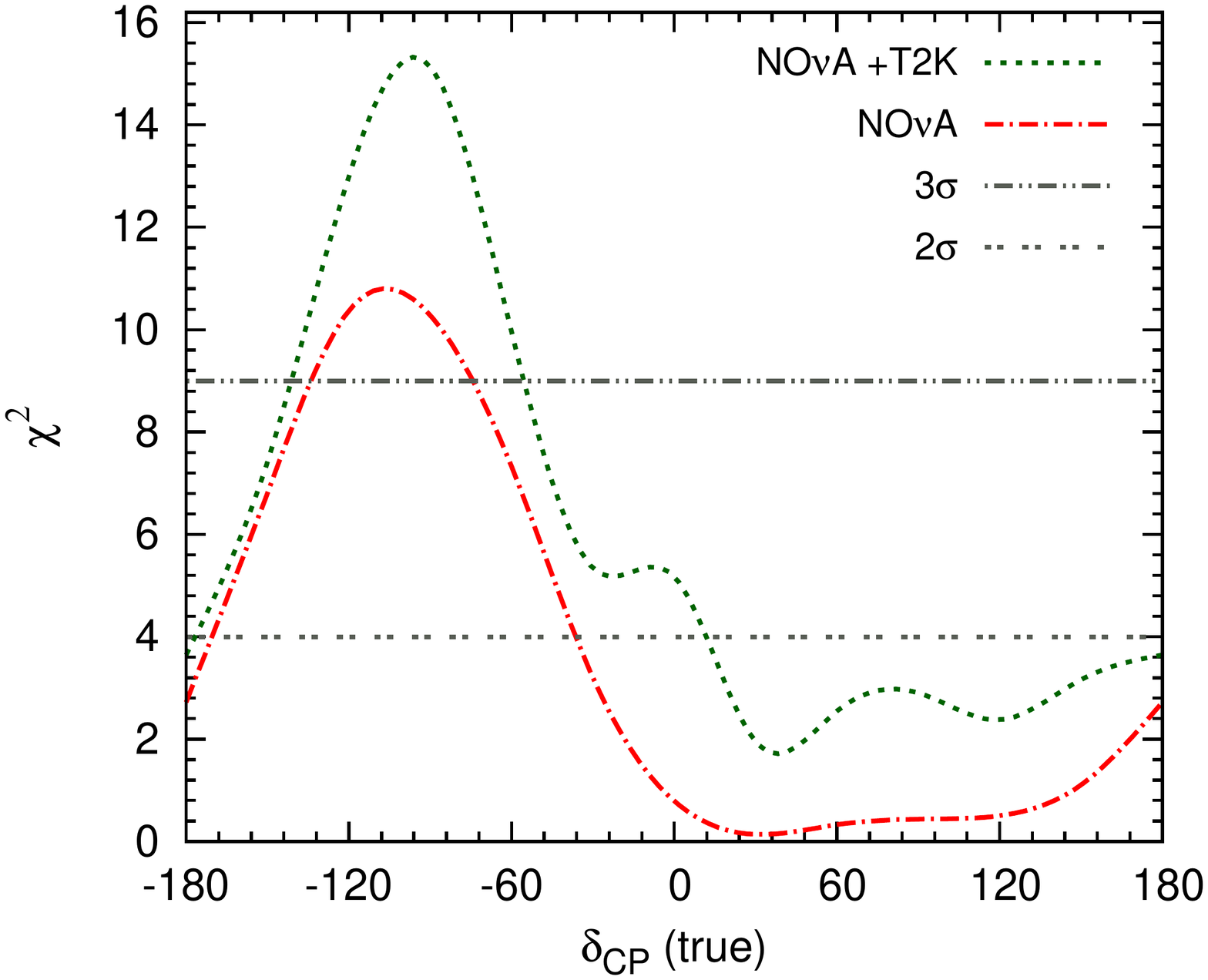}
  \includegraphics[height=5cm,width=5cm]{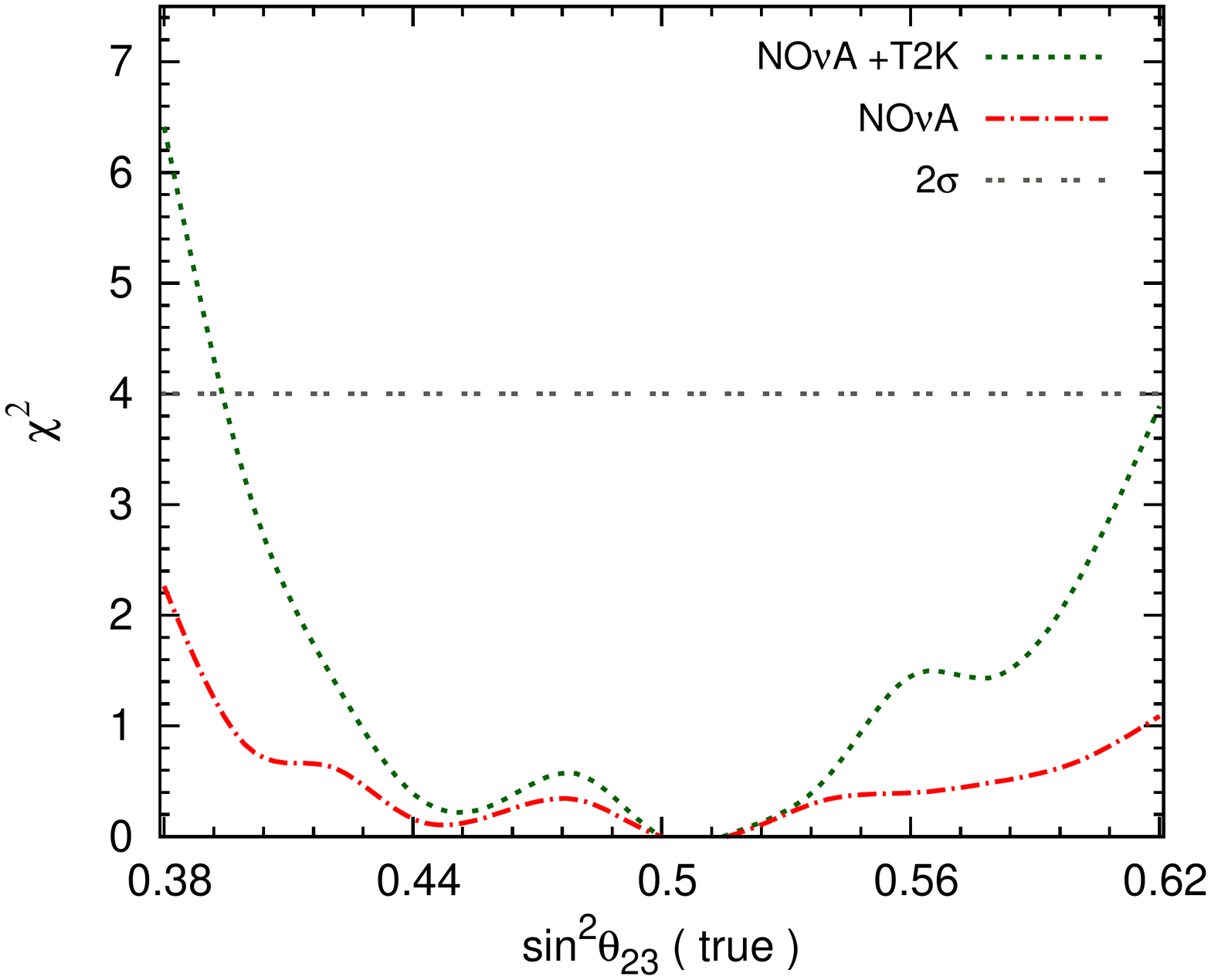}
  \includegraphics[height=5cm,width=5cm]{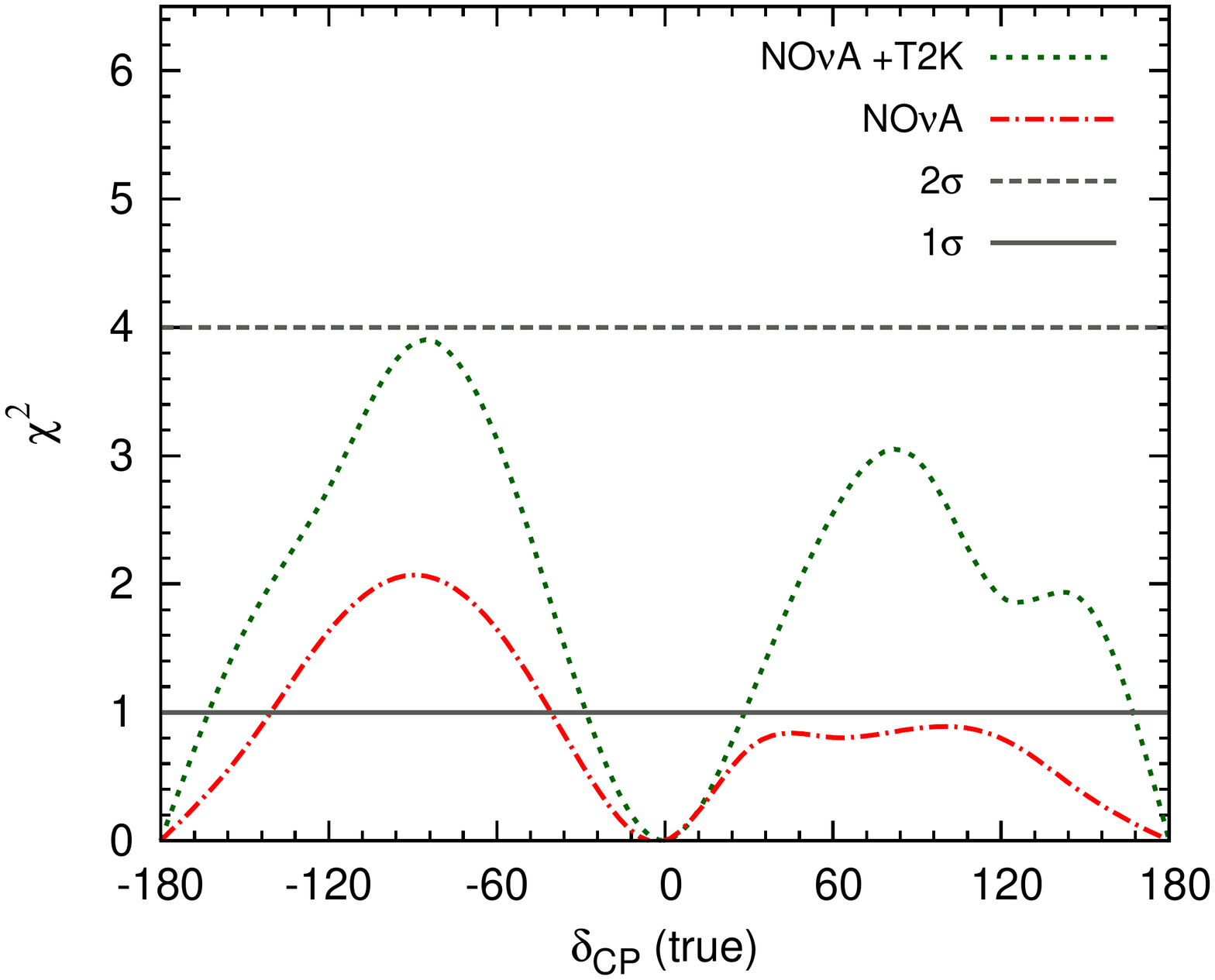}
 \caption{{\label{non-novat2k-sens}} The mass hierarchy, octant, and CPV sensitivities for NO$\nu$A and combination of T2K  and NO$\nu$A are respectively given in left, middle, and right panels. Neutrino mass hierarchy is assumed to be normal and the values of non-unitarity parameters are $\eta_{21}=0.026$ and $\phi_{21}=0$.}
 \end{center}
 \end{figure}

The determination of unknowns in neutrino sector is a challenging task due to the existence of four-fold degeneracies among the oscillation parameters in the standard neutrino oscillation framework. The degeneracy due to sign of $\Delta m_{31}^2$ is known as hierarchy degeneracy and the degeneracy in which one cann't distinguish between $\theta_{23}$ and $(\pi/2 -\theta_{23})$ is known as octant degeneracy. One of the best ways to show these degeneracies is by looking at allowed parameter space in  $\sin^2\theta_{23}-\delta_{CP}$ plane for four different cases i.e., Normal Hierarchy-Higher Octant (NH-HO), Normal Hierarchy-Lower Octant (NH-LO), Inverted Hierarchy-Higher Octant (IH-HO), and Inverted Hierarchy-Lower Octant (IH-LO).\\
In order to obtain the allowed parameter space for NH-HO for three flavor oscillation framework, we assume that neutrino mass hierarchy to be normal and  $\theta_{23}$ to lie in the higher octant with $\sin^2\theta_{23}=0.56$, and allow the test values of $\delta_{CP}$ and $\sin^2\theta_{23}$ to vary in their allowed parameter range. Finally we obtain the minimum $\chi^2$ by doing marginalization over $\Delta m_{31}^2$. It should be noted that for non-unitary, case we assume the true values of non-unitary parameters $\eta_{21}=0.026$ and $\phi_{21}=0$ and while finding minimum $\chi^2$, we also do marginalization over $\phi_{21}$. We repeat the same for NH-LO case wherein we assume the true value of $\theta_{23}$ to lie in the lower octant with $\sin^2\theta_{23}=0.44$. Finally, we repeat the same for inverted hierarchy to get allowed parameter space for IH-HO and IH-LO cases.\\  
In Fig. \ref{non-nova-allowed}, we show the impact of non-unitary mixing on the allowed parameter  space $\sin^2\theta_{23}-\delta_{CP}$. It can be seen from the figure  that in the presence of non-unitary mixing, the  $\sin^2\theta_{23}-\delta_{CP}$ parameter space got enlarged, which means the degeneracy discrimination capability of NO$\nu$A is reduced significantly. In order to know how well the synergy between the T2K and NO$\nu$A helps to resolve the degeneracies among the oscillation parameters in presence of non-unitary mixing, we add T2K data. The experimental configuration of T2K is taken from \cite{Huber:2002mx,Itow:2001ee,Ishitsuka:2005qi}. The results for the synergy between T2K and NO$\nu$A is given in  Fig.\ref{non-novat2k-allowed}. It can be seen from the figure that with the inclusion  of T2K data, the parameter space is reduced and hence, it  improves the degeneracy resolution capability in presence of non-unitary mixing.   Further, we show how the improvement in degeneracy resolution in presence of non-unitary mixing for the synergy of T2K and NO$\nu$A  can affect the mass hierarchy, octant, and CPV sensitivities in Fig.\ref{non-novat2k-sens}. From the figure, it can be seen that while adding the T2K data, there is a significant enhancement in the sensitivities of the unknowns.

\section{Summary and Conclusions}
 In general, the determination of the oscillation parameters are done by taking the assumption that the neutrino mixing matrix is unitary. However, many extensions of the Standard Model require additional fermion fields to incorporate massive neutrino and lead to active-sterile neutrino mixing, 
which gives rise to unitarity violation in active neutrino mixing. In general, the low-scale seesaw models, the so-called  inverse seesaw model,   permits significantly large mixing between the active and sterile neutrinos and gives rise to significant non-unitary lepton mixing.  The constraints on the non-unitarity parameters can be obtained from  the lepton flavor violating decays ($l_i \to l_j \gamma$) which are mediated by the heavy particles present in the model. We have used these constraints on non-unitarity parameters and investigated whether it is possible to probe such non-unitarity parameters at long-baseline experiments. We found that  non-unitarity parameters are sensitive to NO$\nu$A experiment. However, the parameter space allowed by NO$\nu$A experiment at 1$\sigma$ C.L. is $|\eta_{21}|< 0.033$, which is a weaker constrain on this parameter while comparing with the constraint obtained in other physics searches. Therefore, NO$\nu$A experiment is not expected to improve the current knowledge of non-unitarity parameter $\eta_{21}$.

We have also illustrated the impact of non-unitary lepton mixing on the determination of neutrino mass hierarchy, octant of atmospheric mixing angle and CP violating phase. From our analysis, we found that the non-unitarity parameters in  21 sector play crucial role in $\nu_{\mu} \to \nu_e$ oscillation channel.  We also found that non-unitary lepton mixing significantly affect the sensitivities of current unknowns in neutrino sector. In fact, the mass hierarchy sensitivity, octant sensitivity, and CPV sensitivity are deteriorated significantly in presence of non-unitary lepton mixing and the sensitivities are crucially depend up on the new CP-violating phase in the non-unitary mixing. Moreover, the oscillation parameter degeneracy resolution capability of NO$\nu$A experiment is reduced in presence of non-unitarity parameters as they introduced new degeneracies among the oscillation parameters. However, we have seen that the synergy between the currently running experiments T2K and NO$\nu$A has improved degeneracy resolution capability. Therefore, there is a significant enhancement in the sensitivities of unknowns for the synergy of T2K and NO$\nu$A.\\


\newpage
{\bf Acknowledgments}
Authors would like to thank Science and Engineering Research Board (SERB), Government of India for financial support through grant No. SB/S2/HEP-017/2013. SC would like to thank Dr. Sushant K. Raut for many useful discussions regarding MonteCUBES. 

\section*{Appendix}\label{chi}
\subsection*{ A. Details of $\chi^2$ analysis}

The $\chi^2$ analysis is done by comparing true event spectra (predicted event spectra) $N_i^\textrm{true}$ with test event spectra $N_i^\textrm{test}$ (event spectra for alternate hypothesis) and its general form is given by
\begin{equation}
\chi^2_{\rm stat} (\vec{p}_\textrm{true},\vec{p}_\textrm{test})= \sum_{i\in\textrm{bins}} 2\Big [N_i^\textrm{test} - N_i^\textrm{true}-N_i^\textrm{true}\ln\left(\frac{N_i^\textrm{test}}{N_i^\textrm{true}}\right) \Big ],
\end{equation}
where $\vec{p}$ is the array of standard neutrino oscillation parameters. However, while calculating the $\chi^2$ numerically, we also include the systematic errors by using pull method which is done with the help of nuisance systematics parameters as mentioned in the GLoBES manual.  

 Let us assume that $\vec{q}$ is the oscillation parameter in presence of non-unitary neutrino mixing. Then the sensitivity of non-unitarity parameter $\eta_{21}$ can be evaluated as
\begin{equation}
\chi^2(\eta_{21}^\text{test}) = \chi^2_{SO}- \chi^2_{NU}\;,\end{equation}
where  $\chi^2_{SO}= \chi^2(\vec{p}_\textrm{true},\vec{p}_\textrm{test})$,
$\chi^2_{NU}= \chi^2(\vec{p}_\textrm{true},\vec{q}_\textrm{test})$,
Further, the sensitivities of current unknowns in neutrino oscillation is given by
\begin{itemize}
\item MH sensitivity: 
\begin{eqnarray}
\chi^2_\text{MH} &=& \chi^2_\text{NH} -\chi^2_\text{IH}~~~~(\text{for true normal hierarchy})\\
\chi^2_\text{MH} &=& \chi^2_\text{IH} -\chi^2_\text{NH}~~~~(\text{for true inverted hierarchy})
\end{eqnarray}
\item Octant sensitivity: 
\begin{eqnarray}
\chi^2_\text{Octant} &=& \chi^2_\text{HO} -\chi^2_\text{LO}~~~~(\text{for true Higher Octant})\\
\chi^2_\text{Octant} &=& \chi^2_\text{LO} -\chi^2_\text{HO}~~~~(\text{for true Lower Octant})
\end{eqnarray}
\item CPV sensitivity: 
\begin{equation}
\chi^2_\text{CPV} (\delta^\text{true}_{CP}) = \text{Min}[\chi^2(\delta^\text{true}_{CP},\delta^\text{test}_{CP}=0), \chi^2 (\delta^\text{true}_{CP},\delta^\text{test}_{CP}=\pi)]
\end{equation}

\end{itemize}

Further, obtain minimum  $\chi^2_\text{min}$ by doing marginalization over all oscillation parameter spaces. 


\end{document}